\documentclass[floats,floatfix,amssymb,prd,twocolumn,superscriptaddress,nofootinbib]{revtex4-1}

\usepackage[english]{babel}
\usepackage[utf8]{inputenc}
\usepackage{amsmath}
\usepackage{mathbbol}
\usepackage{amssymb}
\usepackage{graphicx,amsfonts}
\usepackage{float}
\usepackage{epsfig}
\usepackage[colorlinks=true,
linkcolor=blue,
urlcolor=blue,
citecolor=blue]{hyperref}
\usepackage{bm}
\usepackage{mathrsfs}
\usepackage{mathtools}
\usepackage{enumerate}
\usepackage{amsthm}
\usepackage{bbm}
\usepackage{comment}
\usepackage{physics}
\usepackage{url}
\usepackage{upgreek}
\usepackage[svgnames]{xcolor}
\usepackage[title]{appendix}

\begin{document}

\title{Greybody factors, reflectionless scattering modes, and echoes\\ of ultracompact horizonless objects}

\author{Romeo Felice Rosato}
\affiliation{Dipartimento di Fisica, Sapienza Università di Roma \& INFN, Sezione di Roma, Piazzale Aldo Moro 5, 00185, Roma, Italy}

\author{Shauvik Biswas}
\affiliation{School of Physical Sciences, Indian Association for the Cultivation of Science, Kolkata 700032, India}

\author{Sumanta Chakraborty}
\affiliation{School of Physical Sciences, Indian Association for the Cultivation of Science, Kolkata 700032, India}

\author{Paolo Pani}
\affiliation{Dipartimento di Fisica, Sapienza Università di Roma \& INFN, Sezione di Roma, Piazzale Aldo Moro 5, 00185, Roma, Italy}

\begin{abstract}
    Motivated by a recently discovered connection between the greybody factors of black holes and the ringdown signal, we investigate the greybody factors of ultracompact horizonless objects, also elucidating their connection to echoes.
    The greybody factor of ultracompact objects features both low-frequency resonances and high-frequency, quasi-reflectionless scattering modes, which become purely reflectionless in the presence of symmetric cavity potentials, as it might be the case for a wormhole.
    We show that it is these high-frequency (quasi-)reflectionless scattering modes, rather than low-frequency resonances, to be directly responsible for the echoes in the time-domain response of ultracompact objects or of black holes surrounded by matter fields localized at large distances.
\end{abstract}

\maketitle 

\tableofcontents

\section{Introduction}
Despite being a decade-long topic~\cite{Regge:1957td,Chandrasekhar:1985kt}, black-hole~(BH) perturbation theory is recently undergoing a renaissance (see~\cite{ringdownreview} for a recent review). This was mainly driven by the first detection of the ringdown of merger remnants with gravitational waves~(GWs)~\cite{LIGOScientific:2016aoc,LIGOScientific:2020iuh} and by the prospect of BH spectroscopy~\cite{Dreyer:2003bv,Detweiler:1980gk,Berti:2005ys,Gossan:2011ha} to test gravity in the strong-field/highly-dynamical regime~\cite{LIGOScientific:2021sio,Berti:2015itd,Berti:2018vdi,Cardoso:2019rvt}.
The accuracy of these tests will increase in the next few years and will experience a phase transition with next-generation interferometers such as LISA~\cite{LISA:2024hlh}, Einstein Telescope~\cite{ET:2019dnz, Kalogera:2021bya, Hild:2010id,Branchesi:2023mws}, and Cosmic Explorer~\cite{LIGOScientific:2016wof,Essick:2017wyl,Evans:2023euw}. These instruments have the capability of measuring the ringdown with subpercent accuracy~\cite{Berti:2016lat,Bhagwat:2021kwv,Bhagwat:2023jwv}, thus performing unprecedented tests of General Relativity and of the nature of compact objects~\cite{Cardoso:2019rvt}.

The holy grail of BH spectroscopy is the extraction of several quasinormal modes~(QNMs)~\cite{Vishveshwara:1970zz,Kokkotas:1999bd,Berti:2009kk,Konoplya:2011qq} of the merger remnant, as the latter relaxes to a stable end-state.
If the remnant is a BH, General Relativity predicts that the infinite tower of QNMs is uniquely described by its mass and spin, allowing for multiple null-hypothesis tests of gravity~\cite{Isi:2019aib,Franchini:2023eda}, the nature of the remnant~\cite{Maggio:2020jml,Maggio:2021ans,Maggio:2023fwy}, and the astrophysical environment around compact objects~\cite{Barausse:2014tra,Cardoso:2021wlq,Cardoso:2022whc,Destounis:2022obl}.
The need to accurately model the ringdown has added some layers of complexity to the (originally simple) BH spectroscopy program, also leading to interesting and unexpected results such as the impact of environmental effects~\cite{Barausse:2014tra,Barausse:2014pra,Cheung:2021bol,Berti:2022xfj,Biswas:2023ofz,Singha:2023lum} and of different boundary conditions or near-horizon structure~\cite{Cardoso:2016rao,Cardoso:2016oxy,Cardoso:2017cqb,Abedi:2020ujo}, spectral instabilities~\cite{Nollert:1996rf,Daghigh:2020jyk,Jaramillo:2020tuu,Destounis:2021lum,Gasperin:2021kfv,Boyanov:2022ark,Jaramillo:2022kuv,Sarkar:2023rhp,Destounis:2023nmb,Arean:2023ejh,Cownden:2023dam,Destounis:2023ruj,Courty:2023rxk,Boyanov:2023qqf,Cao:2024oud,Cardoso:2024mrw,Ianniccari:2024ysv,Cai:2025irl}, nonlinearities~\cite{Gleiser:1995gx,Gleiser:1998rw,Ioka:2007ak,Nakano:2007cj,Brizuela:2009qd,Pazos:2010xf,Ripley:2020xby,Loutrel:2020wbw,Sberna:2021eui,Cheung:2022rbm,Mitman:2022qdl,Kehagias:2023ctr,Perrone:2023jzq,Cheung:2023vki,Redondo-Yuste:2023ipg,Redondo-Yuste:2023seq,Yi:2024elj,Zhu:2024dyl,Zhu:2024rej}, role of the overtones~\cite{Giesler:2019uxc,Bhagwat:2019dtm,Baibhav:2023clw} and of the late-time tails~\cite{Price:1972pw,Gundlach:1993tp,Barack:1998bw,DeAmicis:2024not,DeAmicis:2024eoy}.

Recently, an approach complementary to the standard QNM-based ringdown tests was proposed using the BH greybody factors~(GFs)~\cite{Oshita:2022pkc,Oshita:2023cjz,Okabayashi:2024qbz}, which are (real) functions of the frequency characterizing the tunnelling probability of perturbations through the BH effective potential~\cite{Hawking:1975vcx}.
GFs were shown to describe the spectral amplitude of the ringdown signal at frequencies higher than that of the fundamental QNM~\cite{Oshita:2022pkc,Oshita:2023cjz,Okabayashi:2024qbz}.
Furthermore, they were shown to be stable under small deformations of the system and to be associated with quantities that can be obtained by a superposition of QNMs, despite the latter being spectrally unstable~\cite{Rosato:2024arw,Oshita:2024fzf}. 

These recent studies have raised several interesting open problems, such as: \emph{What is the underlying reason for the connection between GFs and QNMs?} (see also~\cite{Konoplya:2024lir,Konoplya:2024vuj}) 
\emph{Can we devise tests of gravity with GFs?} 
\emph{Does the GF-ringdown correspondence also extend to horizonless ultracompact objects?~\cite{Cardoso:2019rvt}} and, if so, \emph{What is the connection between GFs and echoes that characterize the ringdown of ultracompact objects other than BHs?}~\cite{Cardoso:2016rao,Cardoso:2016oxy,Cardoso:2017cqb,Abedi:2020ujo}

The scope of this paper is to investigate some of the above questions.
In particular, we study the GFs of various models for horizonless compact objects and compare them with the BH case. As we shall discuss, the GFs display two important features, both associated with the effective-potential cavity felt by perturbations of horizonless ultracompact objects: 
\begin{itemize}
    \item \emph{low-frequency resonances} associated with their long-lived, low-frequency QNMs~\cite{Cardoso:2014sna}, which are quasi-trapped within the cavity~\cite{Macedo:2018yoi};
    \item \emph{high-frequency, quasi-reflectionless scattering modes}, associated with wave interference within the cavity.
\end{itemize}

Interestingly, in the presence of symmetric cavity potentials, as it might be the case for an inter-universe wormhole~\cite{Visser:1995cc}, these latter modes become purely reflectionless.
In any scattering theory, \emph{reflectionless scattering modes}~(RSMs) are special high-frequency, monochromatic waves that do not get reflected by potential barriers in the presence of cavities, due to wave interference~\cite{PhysRevA.102.063511}.
In our context, RSMs correspond to frequencies where the spacetime GF is exactly unity and leave a characteristic pattern in the ringdown spectral amplitude at frequencies higher than the fundamental QNM.
These modes are generically present also in the GW ringdown of ultracompact objects, whenever partial reflection is present near the object (which might occur for exotic compact objects~(ECOs)~\cite{Maggio:2020jml}), horizon-scale structure, or quantum effects~\cite{Chakraborty:2022zlq}).

As we shall discuss, it is these high-frequency (quasi-)RSMs, rather than low-frequency resonances, to be directly responsible for the echoes in the time-domain response of a ultracompact object or of BHs surrounded by matter fields localized at large distances.
To the best of our knowledge, this is the first time RSMs are discussed in a gravitational context and echoes are shown to arise from high-frequency oscillations associated with quasi-RSMs, rather than low-frequency resonances linked to long-lived QNMs.
Finally, we will show that the ringdown spectral amplitude emitted by a point particle infalling toward a horizonless ultracompact object is modulated by the GFs, so the GW signal directly contains high-frequency (quasi-)RSMs.
Henceforth we use $G=c=1$ units.

\section{Greybody factors of ultracompact objects}\label{greyultracompact}

In this section we will analyze the GF and reflectivity (not to be confused with the reflectivity of the object, which will be introduced later on) for ultracompact non-rotating objects, such as wormholes and other models of ECOs~\cite{Cardoso:2019rvt,Maggio:2021ans} (see~\cite{Macedo:2018yoi} for a related study). We further assume that the spacetime geometry outside the ultracompact object (or at least outside an effective radius, $r_0$) is given by the Schwarzschild metric. In both of these cases we are going to analyze the one-dimensional radial wave equation:
\begin{equation}\label{Regge-Zerilli}
\Bigg[{d^2 \over dr_*^2} + \omega^2 - V_l(r)\Bigg] X_{lm\omega}=0\,.
\end{equation}
Here, $r_*$ is the tortoise coordinate, defined by $(dr/dr_*)=1-(2M/r)$ (modulo a constant), where $M$ is the object's mass, $X_{lm\omega}$ is a proxy for the radial part of the (scalar, electromagnetic, or gravitational) perturbations, and $V_l$ is the corresponding effective potential. 
Since this is a second order differential equation, it requires two boundary conditions, one at infinity and the other at inner boundary, which depends on the object under consideration.
Since we are interested in wave scattering, the boundary condition at infinity reads
\begin{equation}\label{boundary_refl/trasm}
{X}_{lm\omega} \to A^{\rm in}_{lm\omega} e^{-i\omega r_*}+ A^{\rm out}_{lm\omega} e^{+i\omega r_*}\,; 
\quad 
r_*\to+\infty
\end{equation}
where we have assumed the time dependence of the perturbation to be $e^{-i\omega t}$, so the first (second) term represents an ingoing (outgoing) wave. The above boundary condition allows us to define important quantities in the study of a wave scattering process. Given a plane wave scattered from infinity, we can introduce its \emph{reflection amplitude}\footnote{We will refer to transmission and reflection amplitudes for the \emph{complex} quantities, whereas the \emph{transmittivity} and \emph{reflectivity} are defined as their absolute value squared.} due to the effective potential barrier, as,
\begin{equation}\label{refl/transmAmplitudes}
R_{lm}(\omega)= {  A^{\rm out}_{lm\omega} \over A^{\rm in}_{lm\omega}}\,.
\end{equation}
This is a complex-valued quantity that represents the relative amplitude of the wave reflected by the potential. Furthermore, it is possible to define the probabilities of the wave being transmitted or reflected in full generality, which we will refer to as transmittivity and reflectivity, respectively:
\begin{equation}\label{refl/transm}
\mathcal{R}_{lm}(\omega)= \left|{  A^{\rm out}_{lm\omega} \over A^{\rm in}_{lm\omega}}\right|^2\,;\qquad 
\Gamma_{lm}(\omega)= 1-\left|{  A^{\rm out}_{lm\omega} \over A^{\rm in}_{lm\omega}}\right|^2\,.
\end{equation}
The transmittivity $\Gamma_{lm}(\omega)$ is also called GF. Typically the definition $\Gamma_{lm}(\omega)=\left| A^{\rm in}_{lm\omega} \right|^{-2}$ is used for the GFs of BHs. However, as discussed in Appendix~\ref{app:GFSforECOs}, for ECOs this expression does not always represent the probability of a wave being transmitted through the system. Nonetheless, it is important to emphasize that, in the cases of both BHs and wormholes, the two definitions are entirely equivalent. See Appendix~\ref{app:GFSforECOs} for a detailed discussion.

The inner boundary condition depends on the model under consideration, and we will discuss it case by case. Generically, it will be related to both outgoing and ingoing modes at the object's surface, with the amplitude of the outgoing mode related to the object's reflection amplitude.

\begin{figure}[t]
    \centering
    \includegraphics[width=0.48\textwidth]{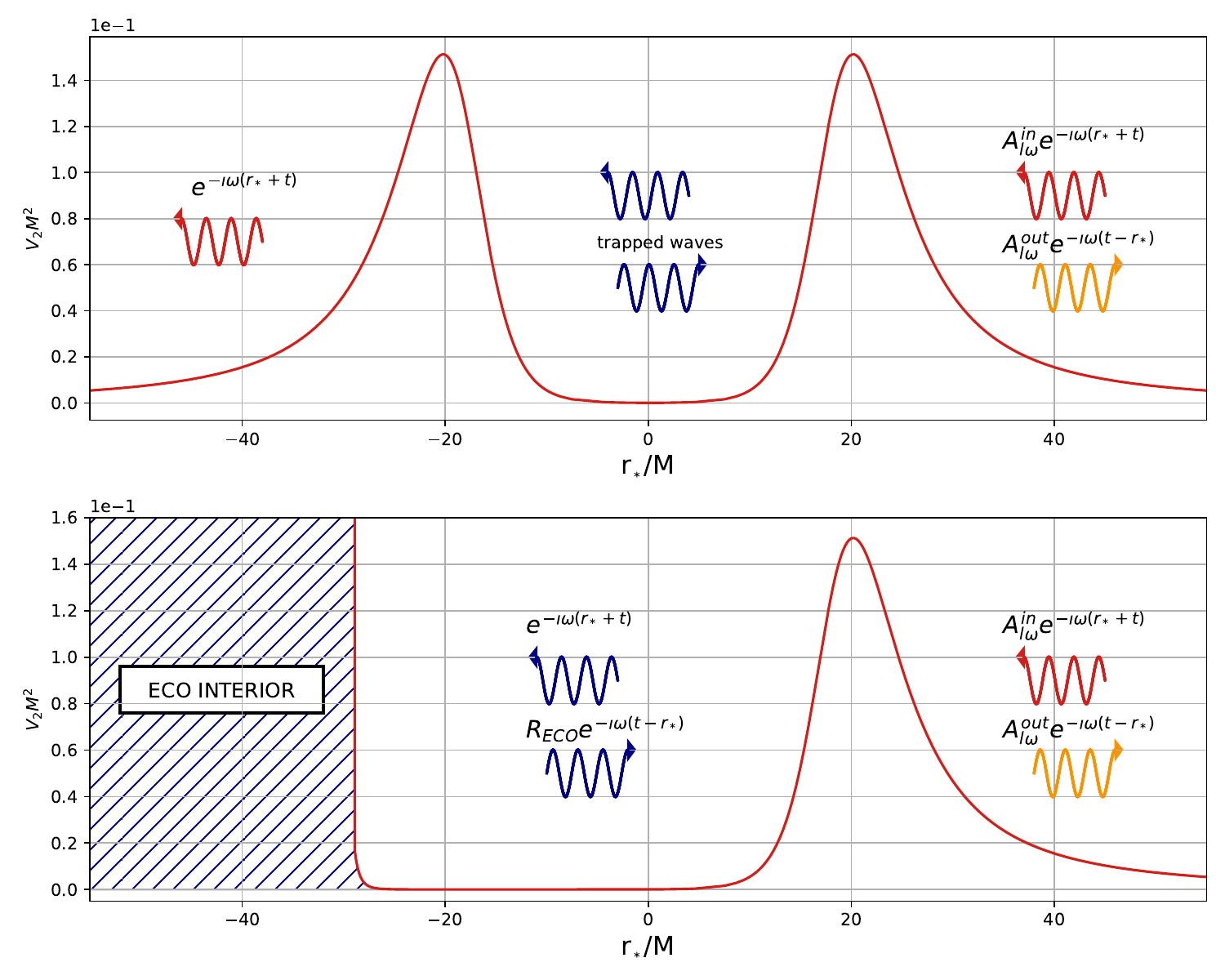}
    \caption{Scattering scheme of plane waves $e^{\pm i\omega r_{*}}$ for a Schwarzschild-like wormhole (top panel) and for a Schwarzschild-like ECO modeled by a reflective potential barrier (bottom panel). See text for further details.}
    \label{schemescattering}
\end{figure}

In the following we shall focus on two classes of ultracompact objects: 
\begin{itemize}
    \item \emph{Schwarzschild-like wormholes}~\cite{Visser:1995cc}, where we consider two Regge-Wheeler\footnote{For concreteness we focus on gravitational axial perturbations, but our analysis is valid in general.} potentials glued at the origin, for which the effective potential reads \cite{Bueno_2018},
\begin{equation}\label{wormholepot}
V_l(r_*)=\theta(r_*)W_l(r_*+r_{*}^0)+\theta(-r_*)W_l(-r_*+r_{*}^0)\,,
\end{equation}
where, $W_l(r_*)$ is the Regge-Wheeler potential; 

\item \emph{Schwarzschild-like ECOs}, whose exterior metric is described by the Schwarzschild metric up to an effective radius $r_0$. For ultracompact objects $r_0/2M\ll 1$.
The ECO interior will be modeled through a reflectivity barrier~\cite{Maggio:2017ivp,Mark:2017dnq,Maggio:2020jml}, with the \emph{surface reflection amplitude} $R_{\rm ECO}(\omega)$ at the ECO radius $r_{0}$ such that, near the effective radius~\cite{Mark:2017dnq,Maggio:2020jml,Chakraborty:2022zlq} 

\begin{equation}\label{ECObcs}
    \Psi\to e^{- {i} \omega (r_*-r_*^0)} + R_{\rm ECO}  e^{ {i} \omega (r_*-r_*^0)}
    \,, \quad r_*\to r_*^0
\end{equation}
where $r_*^0=r_*(r_0)$.
While our formalism is valid for any $R_{\rm ECO}(\omega)$, for concreteness in the following we will consider models with constant $R_{\rm ECO}$ (where the case of a nondissipative object corresponds to $|R_{\rm ECO}|=1$), as well as models featuring Boltzmann reflectivity~\cite{Oshita:2019sat}. 
\end{itemize}
The scattering schemes for Schwarzschild-like wormholes and ECOs are depicted in Fig.~\ref{schemescattering}.

The surface reflection amplitudes analyzed in this work are presented in Table~\ref{tab:ECOs}.
Note that sufficiently compact wormhole spacetimes are included in the reflectivity model by imposing Eq.~\eqref{ECObcs} at the throat and using the surface reflection amplitude given in Table~\ref{tab:ECOs}~\cite{Mark:2017dnq}.

\renewcommand{\arraystretch}{1.5}
\begin{table}[h!]
\centering
\begin{tabular}{|c|c|c|}
\hline
\textbf{ECO Model} & \textbf{Surface reflection amplitude}\\ \hline
Wormhole & $R_{\rm ECO}(\omega)=R_{\rm BH}'(\omega)e^{-2i\omega r_*^0}$  \\ \hline
Constant $R_{\rm ECO}$  &  $R_{\rm ECO}(\omega)={\rm const}$ \\ \hline
Boltzmann $R_{\rm ECO}$ & $R_{\rm ECO}(\omega)=e^{-|\omega|/T_H}$  \\ \hline
\end{tabular}
\caption{Reflection amplitude at the ECO surface for different models. In the constant reflectivity case $R_{\rm ECO}$ is a complex constant (independent of $\omega$) satisfying $\left|R_{\rm ECO}\right|\leq 1$. In the Boltzmann case $T_H$ is the horizon temperature. The wormhole case is described by an effective surface reflection amplitude at the throat~\cite{Mark:2017dnq} (note that the reflectivity $R_{\rm BH}'$ corresponds to the reflection amplitude of the BH potential from the left, see Appendix \ref{app:TransMatrix} for details) only when the two peaks shown in Fig.~\ref{schemescattering} are sufficiently far apart.}
\label{tab:ECOs}
\end{table}

\subsection{Schwarzschild-like wormhole}\label{wormholeGFS}

\begin{figure*}[htbp!]
    \centering
    \includegraphics[width=0.95\textwidth]{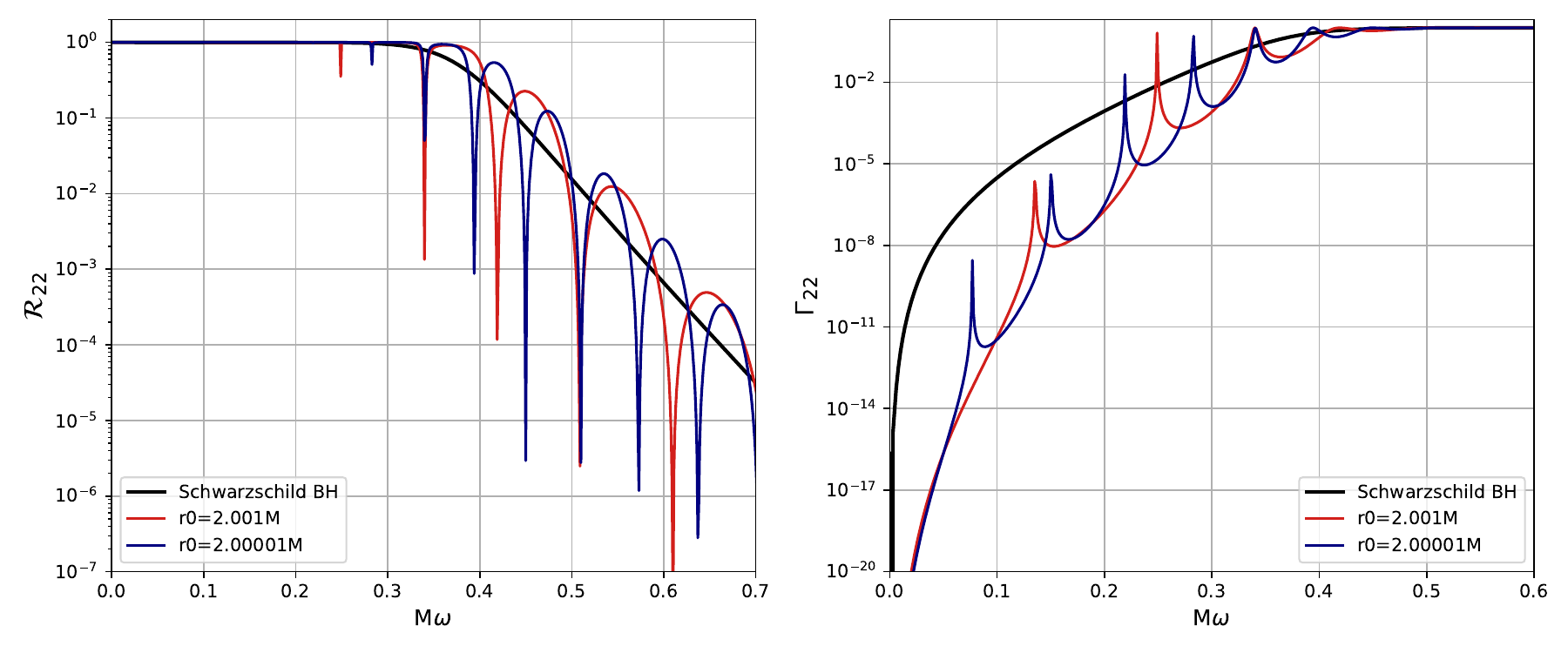}
    \caption{Comparison between reflectivity (left panel) and GFs (right panel) of a Schwarzschild BH (black solid lines) and Schwarzschild-like wormholes for two different values of the throat location $r_0$ (blue and red solid lines).}
    \label{fig:SchwvsWorm}
\end{figure*}

In the case of our Schwarzschild-like wormhole, we require purely left-moving waves at negative infinity, i.e. (see Fig.~\ref{schemescattering}, top panel)
\begin{equation}\label{seconduniverse}
 {X}_{lm\omega} \to  e^{- {i} \omega r_*}\,, \qquad\,\, r_*\to-\infty\,\,.
\end{equation}
 Due to scattering off the double potential barrier, we can define the reflectivity and transmittivity of the background spacetime as in Eq.~\eqref{refl/transm}. Moreover, due to the boundary conditions in the second universe, Eq.~\eqref{seconduniverse}, the GFs can be analogously defined as
\begin{equation}\label{GFSwormholedef}
   \Gamma^{\rm W}_{lm}(\omega) = \frac{1}{\left| A^{\rm in}_{lm\omega} \right|^2}\,,
\end{equation}
as discussed in Appendix~\ref{app:GFSforECOs}.
The physical situation we want to discuss is the following: we are gluing two Schwarzschild geometries at a wormhole throat located at $r_0>2M$ (this is equivalent to using Eq.~\eqref{wormholepot} up to a translation of the origin). We consider the scattering of a plane wave originating from infinity through the double potential barrier. This wave will be partially reflected and partially transmitted with the probabilities defined in Eq.~\eqref{refl/transm}.

In Fig.~\ref{fig:SchwvsWorm} we compare both the $l=2$ GF and reflectivity for a wormhole with those of a Schwarzschild BH, all computed via a direct integration technique~\cite{Pani:2013pma}. In the wormhole case we can notice two interesting features:
\begin{itemize}
    \item Low-frequency resonances, which are more evident in the GF plot (right panel), since the GF is small at small frequencies (see also~\cite{Macedo:2018yoi};
    \item High-frequency resonances and oscillations, which are more evident in the reflectivity plot (left panel), since the reflectivity is small at high frequencies;
    \end{itemize}
Overall, the number of oscillations and resonances increases when approaching the BH compactness, i.e. when $r_0\to 2M$.
That is when the cavity length (throat length) between the two peaks increases.
In the following we will provide an interpretation of these results. 

\subsubsection{Small frequency resonances} Let us focus on the small frequency behavior displayed in Fig.~\ref{fig:SchwvsWorm}. We notice that $\Gamma_{lm}(\omega)$ has a pole whenever  $A^{\rm in}_{lm\omega}=0$ (see Eq.~\eqref{GFSwormholedef}). This would correspond to QNM boundary conditions 
\begin{equation}\label{boundary_QNM}
{X}_{lm\omega}= \begin{cases}
     e^{- {i} \omega r_*} \,\,\,r_*\to-\infty \\
     e^{+ {i} \omega r_*}\,\,\, r_*\to+\infty
 \end{cases}\,.
\end{equation}
This pole does not occur exactly at the location of QNM frequencies, because the QNM frequency is complex, $\omega=\omega_R+i\omega_I$, and $\omega_{I}$ is large for BHs. However, for ultracompact objects the fundamental QNMs have very small imaginary part, corresponding to long-lived modes which are quasi-trapped within the object's photon sphere~\cite{Cardoso:2014sna}. In the wormhole case this happens because of the cavity in between the potential barriers. These long-lived modes give rise to Breit-Weigner resonances on the real axis, corresponding to $\omega=\omega_R$, with a resonance width proportional to $\omega_I$. This is a generic property of weakly-damped harmonic oscillators, featuring long-lived modes (see~\cite{Berti:2009wx} for an example with anti de Sitter BHs, and~\cite{Chandrasekhar:1992ey} in the context of ultracompact stars).

Due to the $Z_2$-reflection symmetry of the wormhole case about the throat, the condition $A^{\rm in}_{lm\omega}=0$ corresponds to either $X_{lm\omega}(r_0)=0$ (Dirichlet modes) or $dX_{lm\omega}/dr_*(r_0)=0$ (Neumann modes). It is easy to see that both corresponds to the boundary condition in Eq.~\eqref{boundary_QNM}.

\begin{figure*}[ht]
    \centering
    \includegraphics[width=0.98\textwidth]{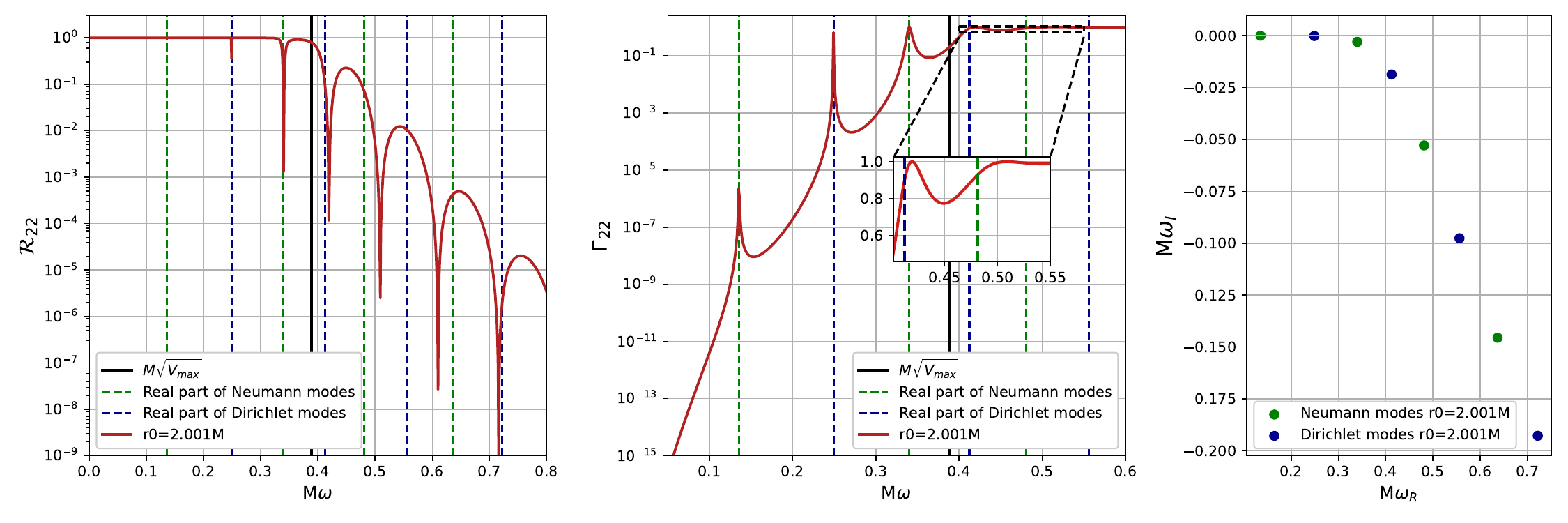}
    \caption{Comparison between the real part of the QNM frequencies and the resonances/oscillations displayed by the GF and reflectivity of a wormhole. As a representative value, we assumed $r_0=2.001M$.
    As a reference, we show the frequency $\omega=V_{\rm max}^{1/2}$ corresponding to the peak of the effective potential, which roughly separates the low-frequency from the high-frequency behavior.    
    }
    \label{fig:QNMsvsresonance}
\end{figure*}

Thus, we expect to see resonances in both the reflectivity and GF for all QNMs with sufficiently small imaginary part, since $A^{\rm in}_{lm\omega_R}\approx0$. This is confirmed in Fig.~\ref{fig:QNMsvsresonance}, where we consider the representative case $r_0=2.001M$ (the result hold also for other values of $r_0\approx 2M$).
It is clear that the low-frequency resonances match the real part of the (long-lived) QNMs. Indeed, from the right panel of the figure one can see that $|\omega_I|\ll\omega_R$, the imaginary part of the first lowest-frequency QNMs is very small, while it grows as $\omega_R$ increases. 

The small-frequency behavior of the GF is discussed in Appendix~\ref{analyticgreybody} by solving the equations analytically in the small-frequency approximation.

\subsubsection{High-frequency oscillations}
Figure~\ref{fig:QNMsvsresonance} also shows that the QNMs do not correspond to the resonances and oscillations displayed at high frequency, in particular above $\omega=\sqrt{V_{\rm max}}$, which roughly separates the low-frequency from the high-frequency behavior.

Therefore, the high-frequency oscillations and resonances must have a different origin. As we shall show, such oscillations are due to a \emph{dephasing} experienced by the plane waves while traveling between the two potential barriers.

This can be better understood via a transfer matrix approach~\cite{Ianniccari:2024ysv}, which we develop in Appendix~\ref{app:TransMatrixWorm} for the wormhole case.
The main idea is to decompose the scattering off the wormhole into two pieces given by the scattering off the individual potentials, using the fact that the transfer matrix is linear~\cite{Ianniccari:2024ysv}.
This can be done exactly when the two potentials are far apart, i.e. when their separation $L\gg M$ (see Appendix~\ref{app:TransMatrix} for details). 
In this limit we obtain the following analytical result for the GF and reflectivity (see Appendix~\ref{app:TransMatrixWorm} for a derivation)
\begin{eqnarray}
\Gamma_{\rm lm}(\omega)&=& \left| {1 \over   \left(\alpha^{\rm in}_{lm \omega}\right)^2e^{-{i} \omega L}-\left(\alpha^{\rm out,*}_{lm\omega}\right)^2 e^{{i} \omega L}}\right|^2\,, \label{GammaTM}\\
\mathcal{R}_{\rm lm}(\omega)&=& \left|\alpha_{lm\omega}^{\rm in}\alpha_{lm\omega}^{\rm out}
-\alpha_{lm\omega}^{\rm in,*}\alpha_{lm\omega}^{\rm out,*}e^{2 i\omega L} \right|^2 \Gamma_{\rm lm}(\omega)\,,\nonumber\\
\label{ReflTM}
\end{eqnarray}
where $\alpha^{\rm out /in}_{lm\omega}$ are the outgoing/ingoing amplitudes in the Schwarzschild BH case for purely ingoing waves at the other universe, i.e., from $r_*=-\infty$. As shown in Appendix~\ref{app:TransMatrixWorm}, Eq.~\eqref{GammaTM} provides an excellent approximation of the exact result when $L\gg M$ and considering a nontrivial dependence of $L=L(\omega)$ to account for the fact that each frequency sees a different effective distance between the potential barriers.

Since $\alpha_{lm\omega}^{\rm in,*}\alpha_{lm\omega}^{\rm out,*}=\alpha_{lm\omega}^{\rm in}\alpha_{lm\omega}^{\rm out}e^{{i}\phi}$, with $\phi=\phi(\omega)$, a frequency-dependent phase, we can recast Eq.~\eqref{ReflTM} as follows
\begin{equation}
\mathcal{R}_{\rm lm}(\omega)= \left|1-e^{{i}\phi(\omega)}e^{2{i}\omega L} \right|^2 \left|\alpha_{lm\omega}^{\rm in}\alpha_{lm\omega}^{\rm out}\right|^2\Gamma_{\rm lm}(\omega)\,.
\label{ReflTMnew}
\end{equation}
The latter explicitly illustrates the key features of the wormhole reflectivity. Specifically, for $\omega > \sqrt{V_{\rm max}}$, we find that $\Gamma_{\rm lm}(\omega)$ rapidly approaches unity while $\alpha_{lm\omega}^{\rm in}\alpha_{lm\omega}^{\rm out}$ vanishes.
Furthermore, one can numerically check that $\phi$ is a slowly-varying function of $\omega$. Consequently, the oscillations of $\mathcal{R}_{\rm lm}(\omega)$ are governed by the term $e^{2i\omega L}$ in the above equation. The period of these oscillations is $\sim \pi / L$. Finally, as will be discussed in the next subsection, the first term in Eq.~\eqref{ReflTMnew} also explains the existence of RSMs, where ${\cal R}_{lm}=0$.

\subsection{RSMs of wormholes}

\begin{figure}[t]
    \centering
    \includegraphics[width=0.52\textwidth]{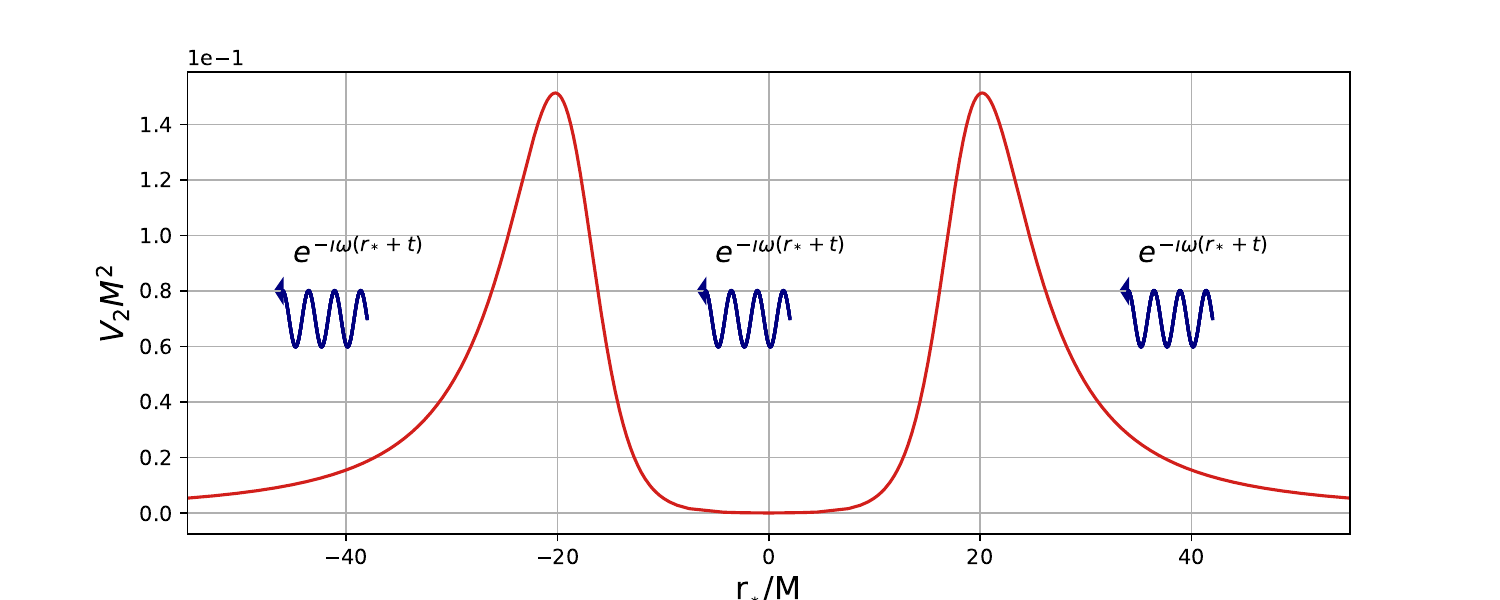}
    \caption{Schematic representation of a RSM for a Schwarzschild-like wormhole. For special discrete (real) frequencies, a plane wave $e^{-i\omega r_{*}}$ propagates from one universe to the other without being altered by the double-potential barrier.}
    \label{schemeRSMs}
\end{figure}

In addition to high-frequency oscillations, Figs.~\ref{fig:SchwvsWorm} and \ref{fig:QNMsvsresonance} show also full-fledged high-frequency resonances where the reflectivity ${\cal R}_{22}=0$, and hence $\Gamma_{22}=1$: the scattering off a wormhole remarkably shows RSMs. These are specific (discrete and countably infinite) frequencies for which waves, upon interacting with a potential, propagate without being reflected (see Fig.~\ref{schemeRSMs} for a schematic representation in the wormhole case).

Here we aim to understand the origin of these special modes. Consider two potential barriers displaced at a distance $L$. Via the transfer matrix formalism (Appendix~\ref{app:TransMatrix}), one can compute the total transmission amplitude $T$ of the system. In the wormhole case, $T$ is the complex amplitude of the wave transmitted to the second universe when a wave of unit amplitude originating from the first universe is scattered off the wormhole. As shown in Appendix~\ref{app:TransMatrixWorm},
\begin{equation}
    {1 \over T}=-{R'_2 \over T_2}{R_1 \over T_1} e^{{i} \omega L}+{1 \over T_1 T_2}e^{-{i} \omega L}\,,
\end{equation}
 where $R_1,R_2$ ($T_1,T_2$) are the reflection (transmission) amplitudes for waves originating from the right of the first and second potentials, respectively, whereas $R'_2$ is the reflection amplitude for waves originating from the left of the second potential. The RSMs of the system are defined such that $\Gamma=|T|^2=1$. This yields the condition
\begin{equation}
    \left|{R'_2 \over T_2}{R_1 \over T_1} e^{{i} \omega L}-{1 \over T_1 T_2}e^{-{i} \omega L}\right|^2=1\,,
\end{equation}
 hence
\begin{equation}
    \left|R'_2 R_1 e^{2 {i} \omega L}-1\right|^2=\left|T_1 T_2\right|^2=\Gamma_1 \Gamma_2 \,.
\end{equation}
 Defining $\mathcal{R}=|R|^2=1-\Gamma$, this implies
\begin{equation}
   1 +\mathcal{R}'_2 \mathcal{R}_1-2 {\rm Re}[R'_2 R_1 e^{2 {i} \omega L}]=1-\mathcal{R}_1 -\mathcal{R}_2 + \mathcal{R}_1\mathcal{R}_2\,.
\end{equation}
 For simplicity, we can restrict to potentials such that reflectivities   from the left and from the right coincide, $\mathcal{R}=\mathcal{R}'$, (which is the case for the Regge-Wheeler, Zerilli and Pöschl–Teller potentials, and in general for all potentials admitting plane waves near the boundaries~\cite{Chandrasekhar:1985kt}). It follows that the condition for RSMs reads
\begin{equation}
\mathcal{R}_1 +\mathcal{R}_2-2 {\rm Re}[R'_2 R_1 e^{2 {i} \omega L}]=\left|R_1-R'_2 e^{2 {i} \omega L}\right|^2=0\,,
\end{equation}
which is satisfied \emph{if and only if}
\begin{equation}
    R_1=R'_2 e^{2 {i} \omega L}\,. \label{RSMcondition}
\end{equation}
Since $R_1=R_1(\omega)$ and $R_2'=R'_2(\omega)$, the above equation might admit a solution for certain $\omega$. However, note that it is a complex equation, which in general does not admit real solutions.
We can derive a \emph{necessary} condition to have RSMs by taking the absolute value squared of it\footnote{A particular case of this general result is when the scattering problem has a parity symmetry~\cite{PhysRevA.102.063511}.}
\begin{equation}
    \mathcal{R}_1=\mathcal{R}_2\,.
\end{equation}
This condition has to be satisfied by RSM frequencies, if the latter exist.
Thus,
\begin{itemize}
     \item whenever $ \mathcal{R}_1\neq \mathcal{R}_2$ for any $\omega$, RSMs do not exist. This can happen, for example, if the two barriers have the same shape but one is rescaled by a multiplicative factor different from unity.\footnote{This can be understood by the fact that the rescaling of the potential corresponds to ${\cal R}(\omega)\to{\cal R}(\omega/\sqrt{b})$, where $b<1$ is the rescaling factor. If ${\cal R}$ is monotonous (as in the case of a single peak barrier) the equation $\mathcal{R}_1(\omega)=\mathcal{R}_1(\omega/\sqrt{b})$ does not admit solutions.} 
     
    \item whenever $ \mathcal{R}_1=\mathcal{R}_2$ RSMs \emph{might} exist. For example, this happens (for any $\omega$) if the two barriers have exactly the same shape, as in the wormhole case under consideration.
    
\end{itemize}

\begin{figure}[H]
    \centering
\includegraphics[width=0.5\textwidth]{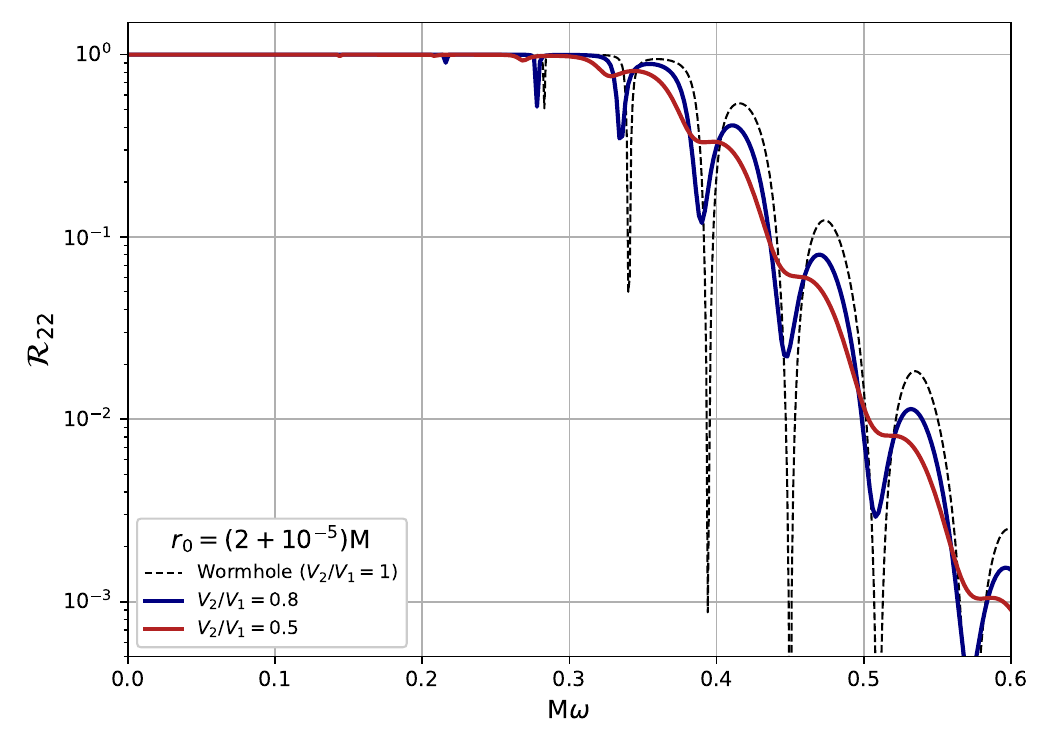}
    \caption{Reflectivity for a double Schwarzschild potential in a framework analogous to the Schwarzschild-like wormhole, but where one of the two potentials is rescaled by a factor $b=V_1/V_2$.
    Unless $b=1$ (wormhole case), RSMs are not present.  
    }
    \label{doubleSchw}
\end{figure}

The previous statements can be checked numerically for the case of a double Schwarzschild barrier. In Fig.~\ref{doubleSchw}, we show a few examples in which one of the barriers is rescaled by a multiplicative factor $b<1$. As expected, while resonances still appear (and their amplitude grows as $b \to 1$), the total reflectivity is never zero unless the barriers are identical.

Let us now discuss in more detail the case of two equal barriers. Since in this case $R_1=\sqrt{\mathcal{R}_1} e^{{i} \phi_1}$ and $R_2'=\sqrt{\mathcal{R}_1} e^{{i} \phi_2}$, the RSM condition~\eqref{RSMcondition} reduces to
\begin{equation}
  e^{{i} \phi_1}=e^{{i} \phi_2 +2{i} \omega L}  \,,
\end{equation}
which is satisfied if and only if 
\begin{equation}
 \phi_1(\omega)= \phi_2(\omega) +2 \omega L+ 2k\pi \,\,\,\,\,k\in \mathbb{Z}\,.
\end{equation}
It follows that RSMs (if they exist) can be identified by solving the previous equation with $k=0$ and for a fixed $L$. In case a solution exists, there will be a countably infinite set of solutions labeled by the integer $k$.

\begin{figure}[H]
    \centering
\includegraphics[width=0.5\textwidth]{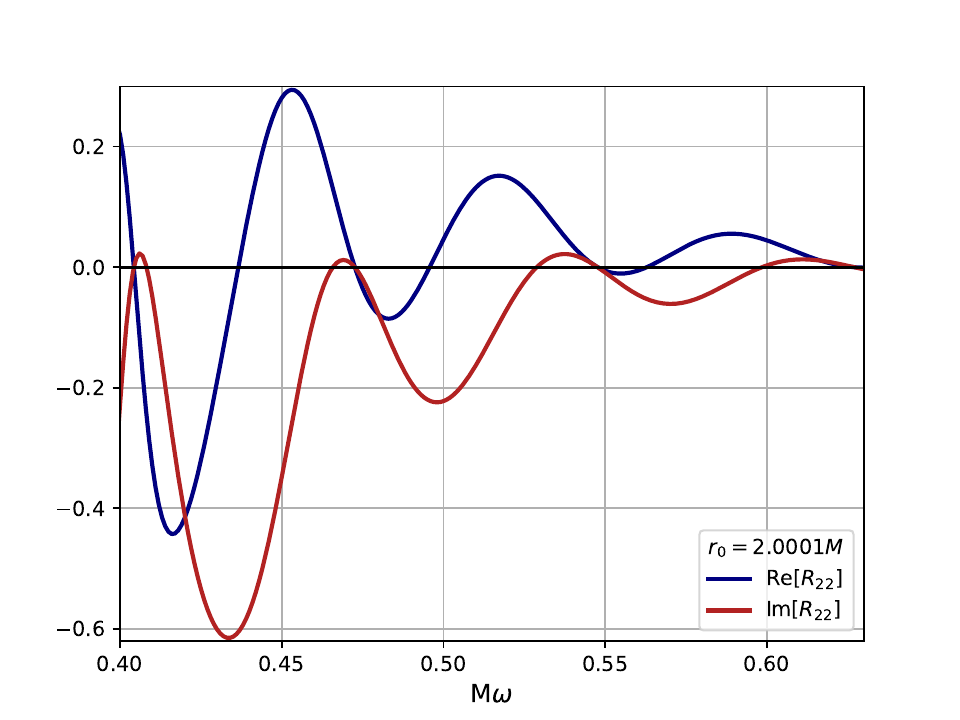}
    \caption{Real and imaginary parts of the reflection amplitude for a $l=m=2$ mode in the Schwarzschild-like wormhole case with $r_0=(2+10^{-4})M$. The solid black horizontal line indicates $0$. It is clear that both the imaginary and real parts of $R_{22}$ vanish for a discrete set of frequencies.}
    \label{RSMS}
\end{figure}

In Fig.~\ref{RSMS}, we present a representative example where RSMs are observed.
This is obtained by solving the full problem for a wormhole, without relying on the transfer-matrix approximation, which is strictly valid only when $r_0\to 2M$ (corresponding to $L\gg M$ in the transfer-matrix model).
As it is evident, for a discrete set of frequencies both the real and imaginary parts of the reflection amplitudes vanish, leading to reflectionless modes. These correspond to the sharp spikes in Figs.~\ref{fig:SchwvsWorm},~\ref{fig:QNMsvsresonance} and \ref{doubleSchw} (for the latter, consider the dashed curve alone), where the vanishing of the reflectivity is not evident due to the logarithmic scale.

\subsection{Schwarzschild-like ECO models}
A star-like ECO can be modeled  by replacing the BH horizon by a partially reflecting surface with frequency dependent reflectivity $R_{\rm ECO}(\omega)$~\cite{Maggio:2021ans,Mark:2017dnq,Maggio:2020jml}. The reflective surface is placed at a location $r_{\rm 0}$ such that $(r_{\rm 0}/2M)-1 \ll 1$, since we are interested in BH-like ECOs, described by a Schwarzschild spacetime in their exterior ($r>r_0$). 
Below, we study the behavior of the reflectivity and transmittivity of the system under axial gravitational perturbations described by Eq.~\eqref{Regge-Zerilli} with the Regge-Wheeler potential. Fig.~\ref{schemescattering} bottom panel shows the schematic representation of the previously studied scattering process in context of star-like ECOs. In the following we are going to analyze two concrete models: with either \emph{Boltzmann} or \emph{constant} surface reflection amplitude (see Table~\ref{tab:ECOs}). 

\subsubsection{Boltzmann ECOs}
It was suggested that quantum effects at the horizon scale can be accounted for by considering an excited multi-level quantum system \cite{Oshita:2019sat}. In this case the effective surface reflection amplitude can be described by the Boltzmann factor, $R_{\rm ECO}=e^{-|\omega|/T_{H}}$, where $T_{H}$ is the temperature of a Schwarzschild BH with same mass of the ECO. 

\begin{figure*}[htbp!]
    \centering
\includegraphics[width=0.95\textwidth]{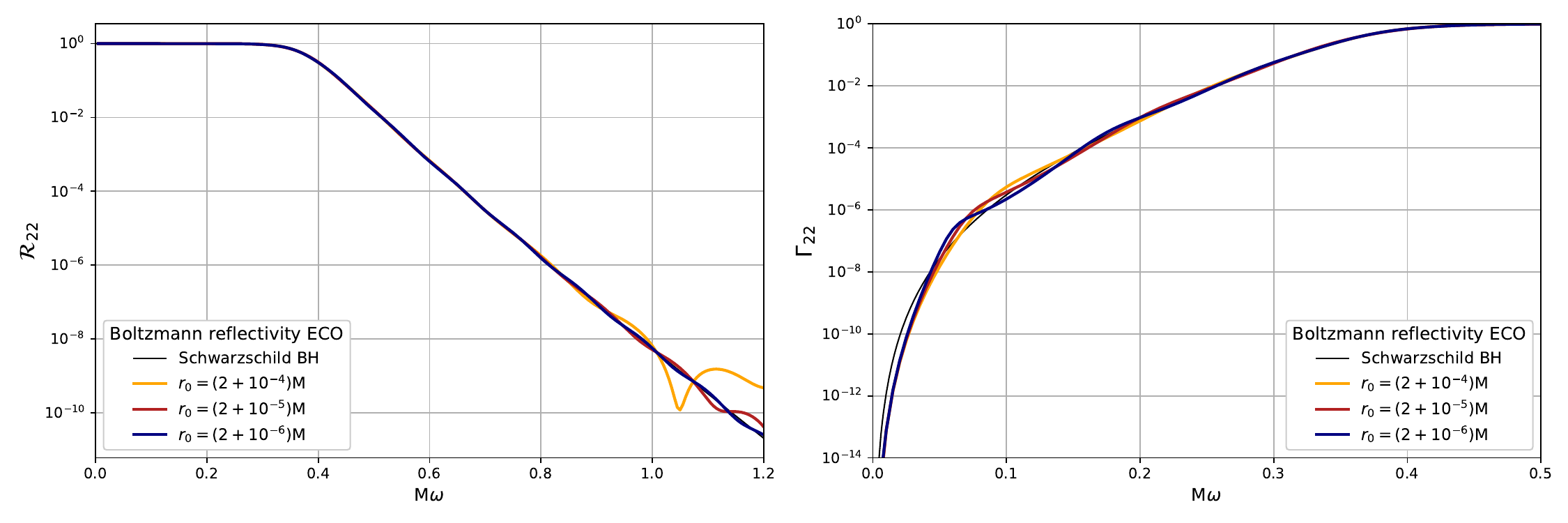}
    \caption{Reflectivity (left panel) and GF (right panel) for an ECO  with Boltzmann reflection amplitude at the surface $r=r_0$, for three choices of $r_0$. Deviations from the BH case (in terms of oscillations) begin to be evident mostly at high frequencies.}
    \label{Boltz}
\end{figure*}

Figure~\ref{Boltz} shows the reflectivity in this model for different values of the ECO radius $r_0$. We can qualitatively notice that for small frequencies the reflectivity is undistinguishable from the standard BH case, while for high frequencies there are evident oscillations. A similar behavior was observed for BH surrounded by matter bumps~\cite{Rosato:2024arw,Oshita:2024fzf}.
For small frequencies some differences are visible also in the GF (right panel). We can understand why this happens by using the transfer matrix formalism (see Appendix~\ref{appendixECOBoltz}), wherein we have computed the GF to be
\begin{equation}\label{reflBoltzECO}
   \Gamma_{\rm lm}^{\rm ECO}(\omega) = { 1-e^{-|\omega|/T_H} \over \abs{\alpha_{\rm lm}^{\rm in}}^2}  \left|e^{-i \omega L}+{\alpha_{\rm lm\omega}^{\rm out,*} \over \alpha_{\rm lm \omega}^{\rm in}}e^{i \omega L}e^{-\abs{\omega}/T_H}\right|^{-2}\,,
\end{equation}
 with $\alpha_{\rm lm\omega}^{\rm in/out}$ the standard Schwarzschild BH ingoing and outgoing amplitudes for plane waves, so that
\begin{equation}\label{reflBoltzECObis}
   \Gamma_{\rm lm}^{\rm ECO}(\omega) = \Gamma_{\rm lm}^{\rm BH}(\omega)  \left(1-e^{-2|\omega|/T_H}\right) \left|1+{\alpha_{\rm lm}^{\rm out,*} \over \alpha_{\rm lm}^{\rm in}}e^{2 i \omega L}e^{-\frac{\abs{\omega}}{T_H}}\right|^{-2}\,.
\end{equation}
 The Boltzmann factor $e^{-|\omega|/T_{H}}$ vanishes exponentially as the frequency increases, while the Schwarzschild reflectivity $1-\Gamma_{lm}^{\rm BH}$ is approximately unity up to $M\omega\approx 0.3$ and then has an exponential fall off. 
 We can then distinguish three frequency regimes:
\begin{itemize}
    \item \emph{Small frequency:} The Boltzmann factor is much smaller than the Schwarzschild reflection amplitude. However, for really small frequencies the correction is not completely negligible, as both quantities are near unity. This can be seen in Fig.~\ref{Boltz} right panel, which shows the GFs for some examples of Boltzmann reflecting ECOs. The dephasing here does not play an important role, being the frequency almost zero. Indeed, as one can see from Eq.~\eqref{reflBoltzECO}, when considering the correction with respect to the BH case, the main role is played by the factor $\left(1-e^{-2|\omega|/T_H}\right)\approx 2|\omega|/T_H$ for small frequencies.
    \item \emph{Intermediate frequency:} In Eq.~\eqref{reflBoltzECObis} the term modulated by the Boltzmann factor can be now neglected, since $\abs{{\alpha_{\rm lm}^{\rm out,*} \over \alpha_{\rm lm}^{\rm in}}e^{-\abs{\omega}/T_H}} \approx e^{-\abs{\omega}/T_H}<1$. Hence the GF (and consequently the reflectivity) starts to be practically indistinguishable from the standard BH case in this regime. This can be seen in both panels of Fig.~\ref{Boltz}. This regime actually lasts until $M\omega \approx 0.6$ since there the Boltzmann factor is still negligible with respect to $\abs{\alpha_{\rm lm}^{\rm out} /\alpha_{\rm lm}^{\rm in}}$.
    \item \emph{High frequency:} For high values of $\omega$ the deviation of the GF with respect to the BH case is still negligible. However, when considering the reflectivity, both the BH reflectivity and the Boltzmann factor have an exponential decay to $0$. Hence the effect of the partially absorbing boundary starts to have a noticeable impact on the final reflectivity. In this high-frequency regime the final result is also affected by the dephasing. This can be seen in Fig.~\ref{Boltz} left panel, where for high values of $M\omega$ the ECO reflectivity has oscillations which are not present in the standard BH case, and depend on $r_0$. Note, however, that the absolute effect is small, since the reflectivity is exponentially suppressed.
\end{itemize}

\subsubsection{ECOs with constant surface reflection amplitude}
\begin{figure*}[htbp!]
    \centering
\includegraphics[width=0.95\textwidth]{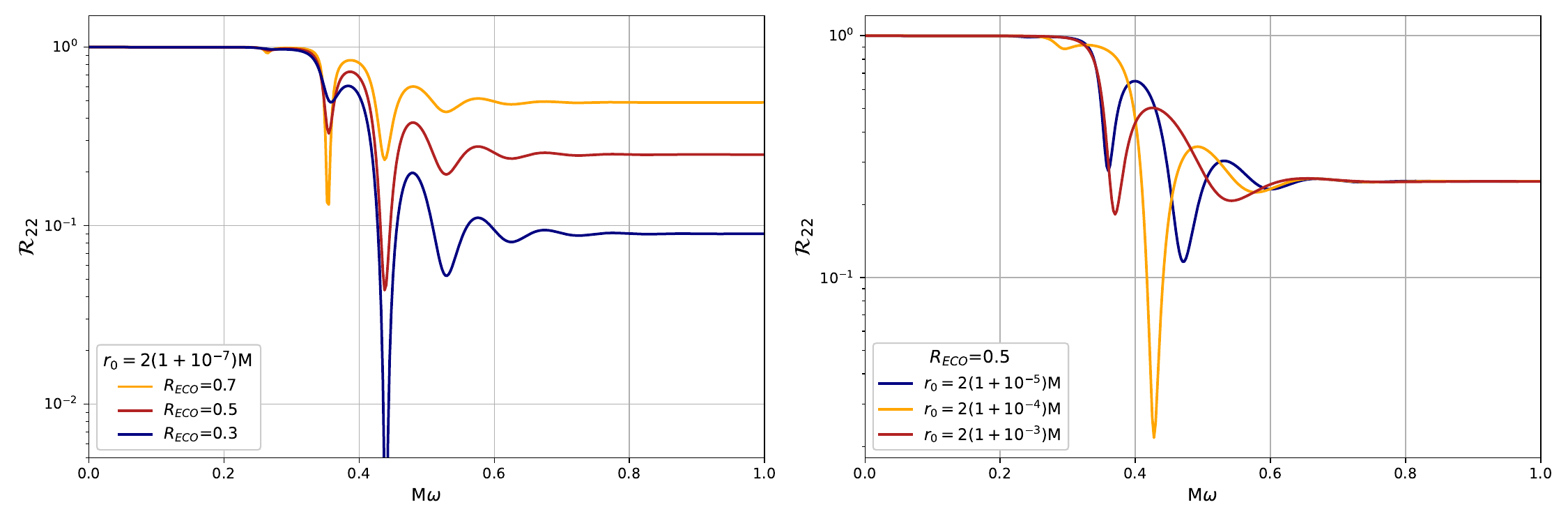}
    \caption{Reflectivity in some ECO models with constant reflectivity at the effective radius. In the left panel we vary $R_{\rm ECO}$ for a fixed value of the radius $r_0=2M(1+10^{-7})$. In the right panel we vary the radius keeping $R_{\rm ECO}=0.5$. Approaching the BH compactness ($r_0\to 2M$) the number of oscillations in $\mathcal{R}_{\rm lm}$ increases.}
    \label{ECO_gfs_constantRefl}
\end{figure*}

It is interesting to analyze the feature of an ECO modeled by a potential barrier at the ECO radius with frequency independent reflectivity $R_{\rm ECO}$. In Fig.~\ref{ECO_gfs_constantRefl} we show the reflectivity of the system for different values of $R_{\rm ECO}$ and of the ECO radius. We recall that  the boundary conditions we are using are
\begin{equation}\label{boundary_ECOconst}
{X}_{lm\omega}= \begin{cases}
     e^{- {i} \omega (r_*-{r_*^0})} + R_{\rm ECO}  e^{ {i} \omega (r_*-{r_*^0})}\,\,\,r_*\to {r_*^0} \\
    A^{\rm in}_{lm\omega} e^{- {i} \omega r_*}+ A^{\rm out}_{lm\omega} e^{+ {i} \omega r_*}\,\,\, r_*\to+\infty
 \end{cases}\,,
\end{equation}
where $r_*^0=r_*(r_0)$ is the location of the effective radius in tortoise coordinates.

From Fig.~\ref{ECO_gfs_constantRefl} we notice evident oscillations in the reflectivity, once again due to the creation of a cavity~\cite{Macedo:2018yoi}. This is discussed in Appendix~\ref{appendixECOconst}. It is interesting to notice that for high frequencies the system reflectivity approaches $\abs{R_{\rm ECO}}^2$. This is expected since for really high frequencies the Schwarzschild effective potential barrier is negligible, hence the waves only experience the potential barrier at the radius $r_0$, which has constant reflectivity for any frequency.


\section{Quasi-RSMs and the origin of echoes}
 In this section we will find a novel connection between the (quasi)-RSMs discussed in the previous section and the origin of echoes appearing in the  time-domain ringdown signal of partially reflective ultracompact objects~\cite{Cardoso:2016rao,Cardoso:2016oxy,Cardoso:2017cqb,Abedi:2020ujo,Dey:2020pth,Biswas:2022wah}.
 For concreteness we will focus on the wormhole case, but the discussion is general.

\begin{figure*}[htbp!]
    \centering
\includegraphics[width=0.95\textwidth]{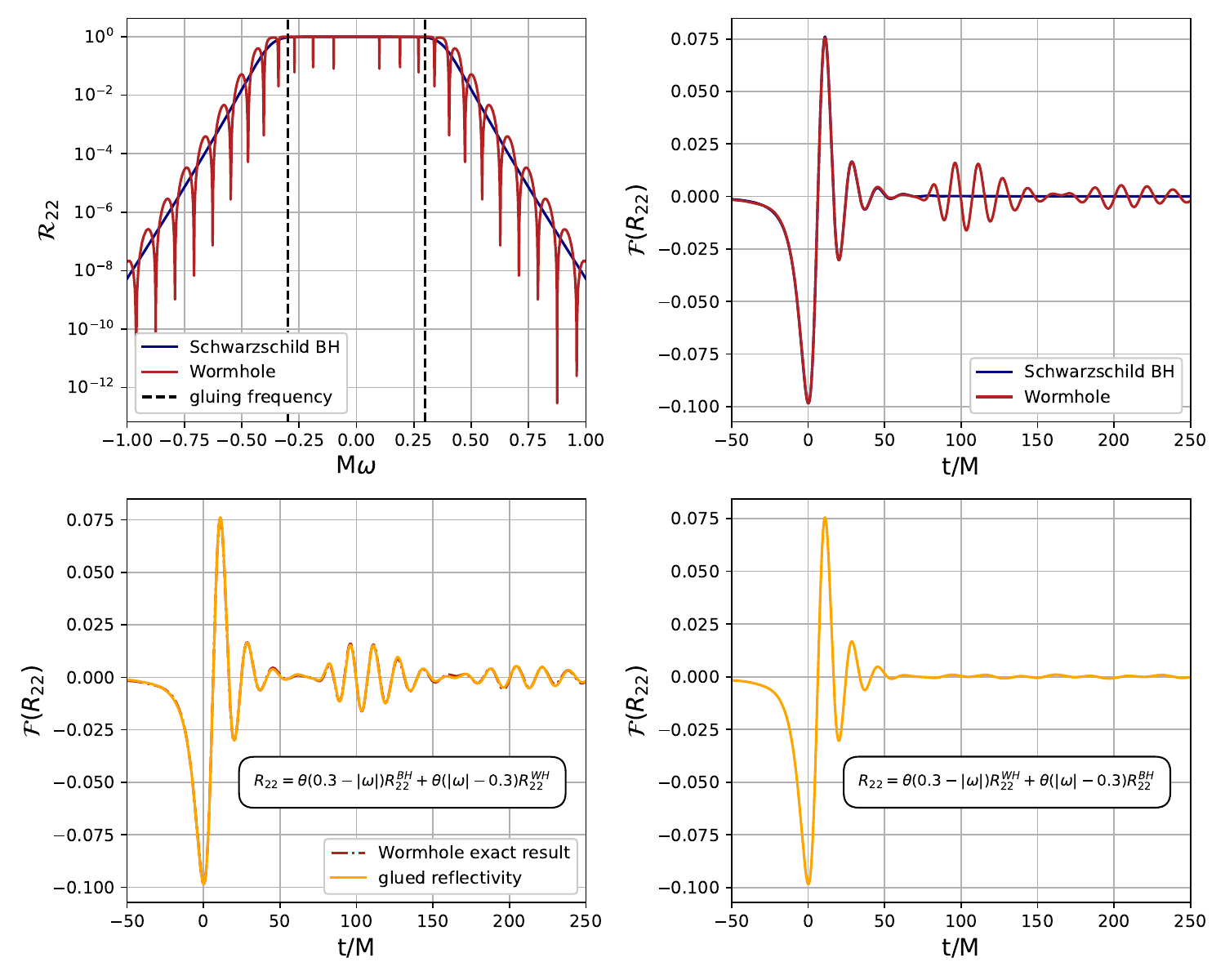}
    \caption{All of the wormhole quantities showed in the plots refer to a wormhole with gluing radius $r_0=2.0001M$. In the left upper panel we show the reflectivity for a Schwarzschild-like wormhole (red solid line) in comparison with the Schwarzschild BH case (blue solid line). In the right upper panel the Fourier transform of Eq.~\eqref{FTrefl} is showed for the two cases. In the wormhole case echoes are present. In the left lower panel the Fourier transform of the model Eq.~\eqref{secondgluing} (solid line) is showed in comparison withe the exact wormhole result (dashed dotted line). The two curves are in perfect agreement, highlighting the origin of echoes from the large frequency behavior. In the right lower panel the Fourier transform of the model Eq.~\eqref{firstgluing} is showed and wormhole echoes are not appearing.}
    \label{fig:echoes}
\end{figure*}

 It is instructing to compute the Fourier transform of the reflection amplitude, as the absolute value of this quantity modulates the spectral amplitude of the emitted GWs during the BH ringdown~\cite{Oshita:2022pkc,Oshita:2023cjz,Okabayashi:2024qbz,Rosato:2024arw}.
The Fourier transform reads
\begin{equation}\label{FTrefl}
    \mathcal{F}(R_{lm})={1 \over 2\pi}\int_{-\infty}^{+\infty}d\omega R_{lm} e^{-{i} \omega t}\,.
\end{equation}
In Fig.~\ref{fig:echoes} we compare $\mathcal{F}(R_{lm})$ in the standard Schwarzschild BH case and in the wormhole case (upper right panel). It is interesting that in the latter case $\mathcal{F}(R_{lm})$ displays echoes analogous to those displayed in the GW signal of a ringing wormhole~\cite{Cardoso:2016rao,Cardoso:2016oxy}.
In the upper left panel we compare the reflectivity of the wormhole with that of the BH, again showing that the differences arise in the form of low-frequency resonances and high-frequency oscillations. Hence, the echoes seen in the time domain must come from either one of these two features.

In order to fully understand the origin of such echoes, we compute $\mathcal{F}(R_{lm})$ in two distinct cases: (i)~neglecting the small frequency oscillations; and (ii)~neglecting the high frequency ones. This is achieved by gluing $R_{22}$ of a Schwarzschild BH ($R^{\rm BH}_{22}$, where no oscillations and resonances are present) to $R_{22}$ for the wormhole ($R^{\rm WH}_{22}$). The gluing is done at $M\omega=0.3$, which roughly separates the low-frequency from the high-frequency regime and is a value where the gluing can be done smoothly. We perform such gluing either by defining:
\begin{equation}\label{firstgluing}
    R_{22}=\theta(|\omega|-0.3)R^{\rm BH}_{22}+\theta(0.3-|\omega|)R^{\rm WH}_{22}\,,
\end{equation}
or
\begin{equation}\label{secondgluing}
    R_{22}=\theta(|\omega|-0.3)R^{\rm WH}_{22}+\theta(0.3-|\omega|)R^{\rm BH}_{22}\,,
\end{equation}
where $\theta(x)$ is the Heaviside function.
The first includes the contribution from the small frequency resonances in the final Fourier transform while keeping the BH response at high frequency. The second case does the opposite: it includes the high-frequency oscillations while assuming the BH response at small frequency. 
These two cases are shown in the bottom panels of Fig.~\ref{fig:echoes}, where we see that the the Fourier transform including only the high frequency oscillations (second case) perfectly reconstructs the exact wormhole curve (bottom left panel case), while the first case reproduces the BH case (right bottom panel).
We can conclude that echoes are completely due to the high-frequency oscillations and that the low-frequency resonances have a negligible role in the prompt ringdown and first echoes.

\begin{figure}[htbp!]
    \centering
\includegraphics[width=0.5\textwidth]{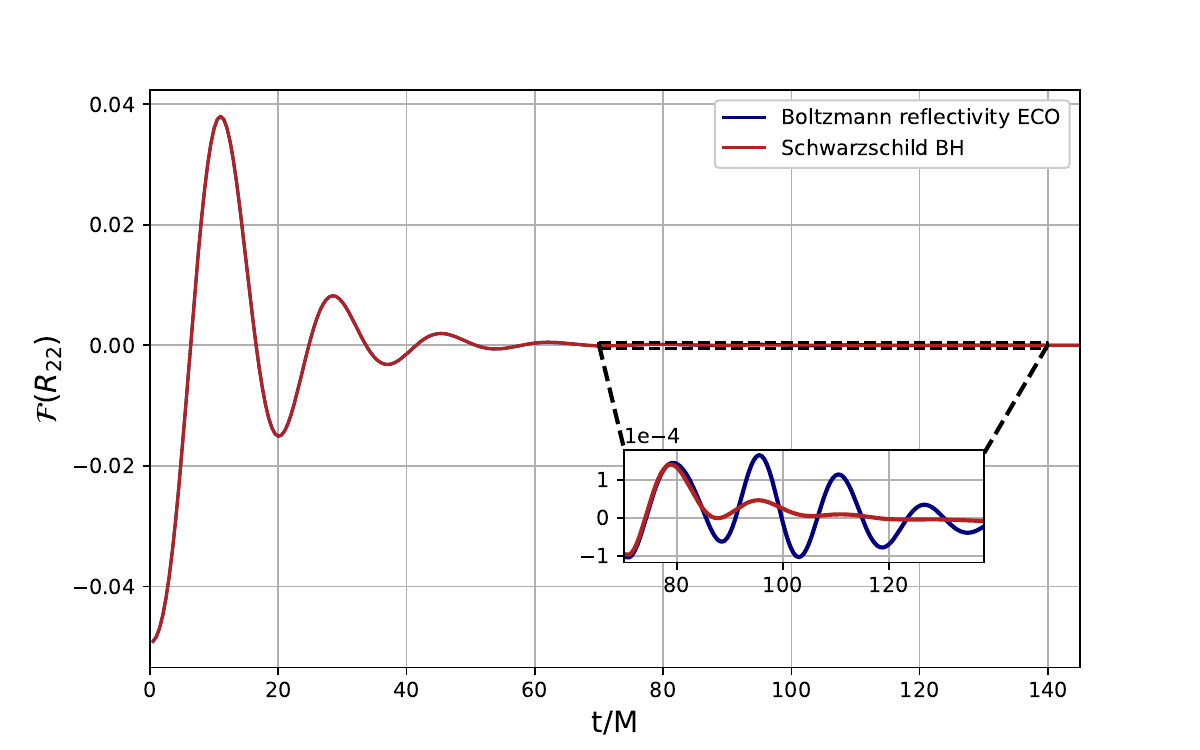}
    \caption{Fourier transform of the complex reflectivity (Eq.~\eqref{FTrefl}) for a Boltzmann reflectivity ECO (blue solid line) compared with the Schwarzschild BH case (red solid line). The ECO radius is $r_0=(2+10^{-9})M$. Notice that the echo appears around $\sim 2 r_*(r_0)$, roughly the length traveled by the plane waves in the cavity.}
    \label{echos_Boltz}
\end{figure}

We verified the previous picture also in the Boltzmann reflectivity ECO case, where echoes are again present in the Fourier transform of the complex reflectivity (see Fig.~\ref{echos_Boltz}). In this case the echo amplitude is very small since the surface reflection amplitude at the relevant (high) frequency is suppressed in the model, and so is the total reflectivity (see Fig.~\ref{Boltz}).

Thus, we showed how the echoes in the ringdown are connected to the high frequency behavior of the GF, rather than the small frequency one. This is due to the well-known fact that the late-time behavior of a Fourier transform is not determined by the low-frequency behavior of the spectrum but by its small-scale variations in frequency. Indeed, at high frequencies, we observe evident oscillations in the reflectivity. These oscillations are related to the phase shift \( e^{2 i \omega L} \) and, consequently, they occur on a scale \( \Delta \omega \sim \pi /L \), impacting the signal at times \( t \sim 2L \) or integer multiples thereof. Notice that also the low-frequency resonances can affect the signal. However, unlike the oscillations appearing at high frequency, these resonances are very narrow (their width $\delta \omega$ is much smaller than $1/L$). Thus, their effect on the time-domain signal becomes significant only at very large times, of the order $1/\delta\omega\gg L$.

\section{GF and GW spectral amplitude}
\begin{figure}[htbp!]
    \centering
\includegraphics[width=0.45\textwidth]{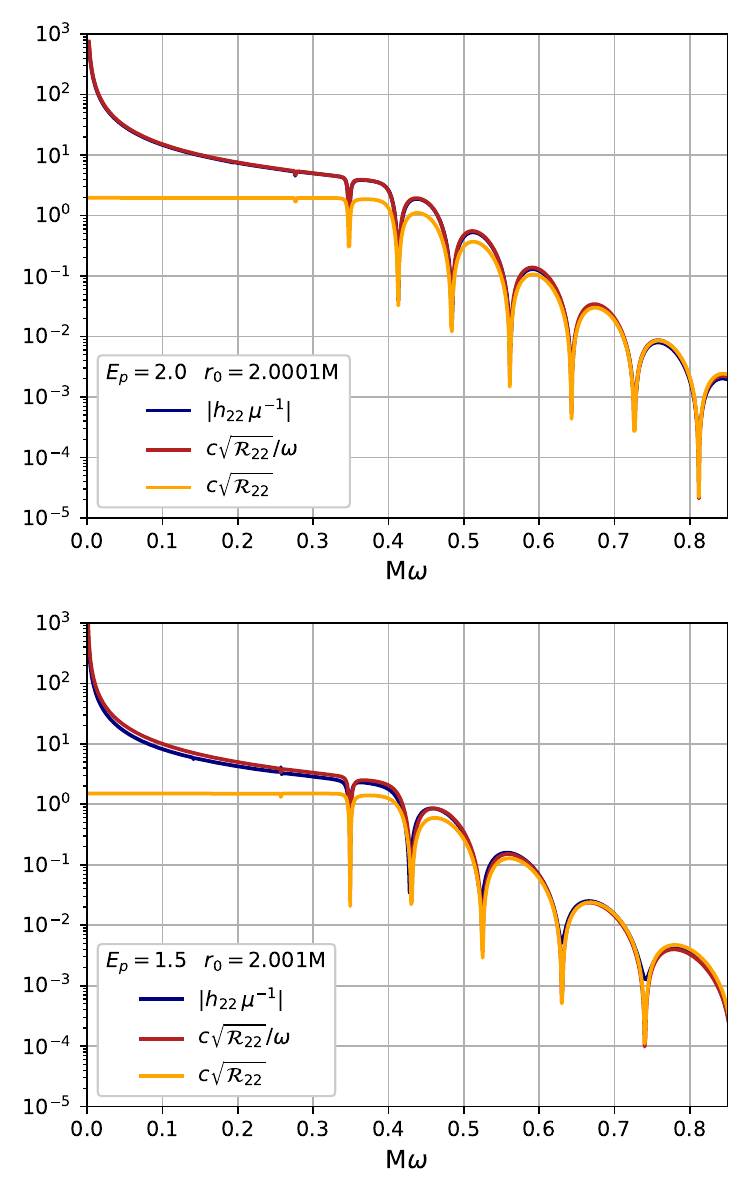}
    \caption{The spectral amplitudes for $l=2$ modes emitted by a point particle with mass $\mu$ in radial infall with specific energy $E_p$ (see legends in each panel). For comparison, we also show two models $\propto\sqrt{\mathcal{R}_{lm}}/\omega$ and $\propto\sqrt{\mathcal{R}_{lm}}$. For all cases with $E_p>1$, the model $\propto\sqrt{\mathcal{R}_{lm}}/\omega$ accurately describe the spectral amplitude at all frequencies.}
    \label{h22wormhole}
\end{figure}

In this section we consider a point particle with mass $\mu$ and specific energy $E_p\geq1$ in radial infall onto a compact object, focusing on the wormhole case for concreteness. The emitted radiation can be described through a Zerilli-like equation with source
\begin{equation}\label{Regge-Zerilli_source}
 \Bigg[{d^2 \over dr_*^2} + \omega^2 - V_l(r)\Bigg] X_{lm\omega}=S_{lm\omega}\,,
\end{equation}
where the source term and the potential are specified, e.g., in \cite{Cardoso:2016rao}. 
In the BH case, the emitted GWs in this toy-model have been showed to be well described by the reflectivity of the homogeneous version of Eq.~\eqref{Regge-Zerilli_source}~\cite{Oshita:2022pkc,Oshita:2023cjz,Okabayashi:2024qbz,Rosato:2024arw}. Here we show that the very same model works for wormholes (and other ECOs).
In Fig.~\ref{h22wormhole} we show the spectral amplitude for some values of $E_p$ and of the gluing point $r_0$. We compute the spectral amplitude as 
\begin{equation}
    h_{22}=h_{22,+}+{i} h_{22,\cross}={e^{{i} \omega r_*} \over 2 {i} \omega A^{\rm in}_{22\omega}} \int_{-\infty}^{+\infty} dr_* X^{\rm hom}_{22\omega}S_{22\omega}\,,
\end{equation}
so that emitted GWs behave as $\propto h_{22}/r$. In both cases showed in the panels the spectral amplitude is well modeled by a \emph{one-parameter} function $\sim \sqrt{\mathcal{R}_{lm\omega}}/\omega$. This is true at all frequencies, in agreement with previous findings in the BH case~\cite{Rosato:2024arw}. This shows that the reflectivity $\mathcal{R}_{lm}=1-\Gamma_{lm}$ is more directly imprinted on the ringdown spectral amplitude than the greybody factors $\Gamma_{lm}$. Therefore, the reflectivity is more useful in the context of BH spectroscopy.
Remarkably, the spectral amplitude displays the same RSMs of the reflectivity. This is important for two main reasons. Firstly, it demonstrates that $h_{22}$ is really modulated by $\mathcal{R}_{22}$, i.e. the dependence is $h_{22}=(\mathcal{R}_{22})^{p} f(\omega)$, where $f(\omega)$ has no poles at finite frequency. It turns out that, for any radially in-falling particle, with $E_p>1$ and $L=0$, we have $p=1/2$ and $f(\omega)=1/\omega$. This behavior is analogous to that observed for radially in-falling particles with  $E_p > 1$ and $L = 0$ in the BH case. Furthermore, numerical verification confirms that also the cases $E_p=1$, $L=0$, as well as $E_p=1$, $L>0$, are consistent with the BH scenario. In particular, for $l=2$ gravitational perturbations in the case $E_p=1$, $L=0$ we have $p=1/2$ and $f(\omega)=1/\sqrt{\omega}$, while for $E_p=1$, $L>0$ we have $p=1/2$ and $f(\omega)=1$ \cite{Rosato:2024arw}.

Secondly, this confirms that there is a direct correspondence between the reflectivity (or, equivalently the GFs) and the ringdown spectral amplitude that goes beyond the BH case~\cite{Oshita:2022pkc,Oshita:2023cjz,Okabayashi:2024qbz,Rosato:2024arw}. Finally, this connection gives us a handle to potentially observe the reflectivity from GW signals. Thus, by analyzing the ringdown spectral amplitude in the frequency domain one could extract information about the nature of a merger remnant.
In a forthcoming work~\cite{Rosato2025}, we will explore these findings and avenues in more detail.

\section{Conclusions}
This paper investigated the GFs, RSMs, and their connection to the ringdown of ultracompact horizonless objects. Using models such as Schwarzschild-like wormholes and partially-absorbing ECOs, we systematically analyzed the scattering properties of these objects and compared them to standard BHs. Our results reveal several intriguing aspects of GFs and their connection to the GW signal, offering new perspectives on ringdown-based tests of compact objects.

Firstly, we have demonstrated that ultracompact objects exhibit both low-frequency resonances, associated with long-lived QNMs, and high-frequency quasi-RSMs. The latter arise due to wave interference within the effective potential cavity and, under certain conditions (e.g., symmetric cavity potentials, such as those in Schwarzschild-like wormholes), these modes become fully reflectionless. Notably, we found that these high-frequency quasi-RSMs, rather than low-frequency resonances, are directly responsible for the echoes observed in the GW ringdown signal of horizonless compact objects. This distinction is crucial, as it identifies the high-frequency regime as the dominant contributor to the characteristic time-domain late-time features in the ringdown of these objects. We have also shown that the standard definition of the GF as simply $|A_{lm\omega}^{\rm in}|^{-2}$ does not hold true for ECOs, rather one needs to define the GF as a function of the reflectivity of the ECO. Our analysis suggests that, for ultra-compact objects, reflectivity is a better (and more direct) observable than the GF.

Secondly, we have also highlighted the role of reflectivity as a bridge between the spectral properties of ultracompact objects and the emitted GW signals. The reflectivity modulate the spectral amplitude of the ringdown signal and are directly linked to the high-frequency quasi-RSMs. This correspondence not only strengthens the theoretical understanding of echoes but also offers a practical framework for observing these features in GW data. By analyzing the spectral oscillations induced by the quasi-RSMs, we showed that echoes are inherently connected to the phase shift introduced by high-frequency scattering. This insight reinforces the importance of high-frequency oscillations in the ringdown spectra of horizonless compact objects.

Our findings have broader implications for GW spectroscopy, as they open up new avenues for testing the nature of compact objects and the validity of General Relativity in the strong-field regime. The study of high-frequency quasi-RSMs and reflectivity (as well as GFs) may provide valuable constraints on the near-horizon structure of compact objects, potentially uncovering signatures of quantum effects or exotic physics beyond the standard BH paradigm.

Future work will focus on extracting quantities such as the reflectivities and the (quasi)-RSMs from actual waveforms and real data, also including the crucial effects of the spin~\cite{Rosato2025}.

Finally, we focused here on horizonless ultracompact objects, but as suggested by previous work~\cite{Rosato:2024arw,Oshita:2024fzf,Ianniccari:2024ysv}, we expect the same phenomenology also for BHs surrounded by localized matter fields~\cite{Barausse:2014tra}, which also give rise to cavities, echoes, and potentially quasi-RSMs at high frequencies.
\begin{acknowledgments}
This work is partially supported by the MUR PRIN Grant 2020KR4KN2 ``String Theory as a bridge between Gauge Theories and Quantum Gravity'', by the FARE programme (GW-NEXT, CUP:~B84I20000100001), and by the INFN TEONGRAV initiative. 
Some numerical computations have been performed at the Vera cluster supported by the Italian Ministry for Research and by Sapienza University of Rome. SB would like to thank INFN Rome for providing support during his visit to Sapienza University of Rome.
Research of SC is supported by MATRICS (MTR/2023/000049) and Core Research (CRG/2023/000934) Grants from SERB, ANRF, Government of India. SC also acknowledges the warm hospitality at the Albert-Einstein Institute, Potsdam, where a part of this work was done. The visit was supported by a Max-Planck-India mobility grant. 
\end{acknowledgments}
\appendix
\section{Transmission probability and GF for ECOs}\label{app:GFSforECOs}

In this section, we analyze the problem of transmission and reflection of a wave, which is scattered off by an ECO and point out subtle issues associated with the definition of GF associated with ECO. The definition of transmission and reflection coefficients follow from the following behaviour of the master function, describing the perturbation, in the asymptotic region,
\begin{equation}\label{BCSapp1}
{X}_{lm\omega} \to A^{\rm in}_{lm\omega} e^{-i\omega r_*}+ A^{\rm out}_{lm\omega} e^{+i\omega r_*}\,; 
\quad 
r_*\to+\infty\,.
\end{equation}
Due to the governing equation for the master function $X_{lm\omega}$, given by Eq.~\eqref{Regge-Zerilli}, the Wronskian ${\cal W}$ of two linearly independent solutions is constant with respect to \(r_*\). It is found that \(\mathcal{W}(X_{lm\omega},X^*_{lm\omega}) = 2i\omega (\left|A^{\rm in}_{lm\omega}\right|^2 - \left|A^{\rm out}_{lm\omega}\right|^2)\) at infinity. Let us now specifically consider an ECO, for which in the near-surface region, the perturbation variable can be expressed in terms of ingoing and outgoing plane waves, such that, 
\begin{equation}\label{BCSapp2}
{X}_{lm\omega}\to e^{- {i} \omega (r_*-r_*^0)} + R_{\rm ECO}  e^{ {i} \omega (r_*-r_*^0)}\,, 
\quad 
r_*\to r_*^0\,.
\end{equation}
Here, $R_{\rm ECO}$ is the reflectivity of the ECO. Consequently, in this region the Wronskian associated with the master function reads, \(\mathcal{W}(X_{lm\omega},X^*_{lm\omega}) = 2i\omega (1 - \left|R_{\rm ECO}\right|^2)\). Since Wronskian is conserved, which effectively follows from energy conservation, 
we obtain the following relation:
\begin{equation}\label{ECOenergyconservation}
 \left|A^{\rm in}_{lm\omega}\right|^2 - \left|A^{\rm out}_{lm\omega}\right|^2=  1-\left|R_{\rm ECO}\right|^2\,.
\end{equation}
This differs from the BH and the wormhole cases, where the master function behaves as ${X}_{lm\omega}\to e^{- {i} \omega r_*}$ for $ r_*\to -\infty$, yielding $\left|A^{\rm in}_{lm\omega}\right|^2 - \left|A^{\rm out}_{lm\omega}\right|^2=1$\,. This is different from the corresponding result for ECO, as presented in Eq.~\eqref{ECOenergyconservation}. Notice that this difference does not affect the definition of the reflection amplitude and reflectivity, with the reflection amplitude taking the following expression, 
\begin{equation}
R_{lm}(\omega)={A^{\rm out}_{lm\omega} \over A^{\rm in}_{lm\omega}}\,.
\end{equation}
However, Eq.~\eqref{ECOenergyconservation} forces the GFs definition to be modified  with respect to the BH and wormhole cases, as long as the GF represents the probability of the wave to be transmitted through the system. In particular, we have
\begin{align}
\mathcal{R}^{\rm E}_{lm}(\omega)&=\left|{A^{\rm out}_{lm\omega} \over A^{\rm in}_{lm\omega}}\right|^2 \,,
\label{ECOref}
\\
\Gamma^{\rm E}_{lm}(\omega)&=\left|{1 \over A^{\rm in}_{lm\omega}}\right|^2\left(1-\left|R_{\rm ECO}\right|^2\right)\,,
\label{ECOgfs}
\end{align}
which clearly reduces to the ordinary BH case when $R_{\rm ECO}=0$. 
Note that also the wormhole case when the two barriers are very far apart is included in the above result.
In that case,
$R_{\rm ECO}=R'_{\rm BH} e^{-2i\omega r_*^0}$~\cite{Mark:2017dnq} (see Table~\ref{tab:ECOs}) and one obtains $\Gamma_{lm}^{\rm wormhole}=1/|A^{\rm in}_{lm\omega}|^2$ after rescaling the normalization of the wave at $r_*\to-\infty$ as $e^{-i\omega r_*}\to T_{\rm BH} e^{-i\omega r_*}$, to account for the different waveform normalization in the wormhole and ECO cases.

The previous calculation actually has a rather intuitive physical interpretation. Indeed, if a wave with unit amplitude is generated at infinity, an incoming wave with amplitude $(1/A^{\rm in}_{lm\omega})$, modulo phase terms, reaches the surface of the ECO (as it follows by comparing Eqs.~\eqref{BCSapp1} and~\eqref{BCSapp2}, presented above). However, the surface reflects the wave with a probability $\left|R_{\rm ECO}\right|^2$, and consequently, the probability of the wave of being transmitted inside the object (effectively absorbed) is $(1-\left|R_{\rm ECO}\right|^2)$. Thus, the wave with unit amplitude generated at infinity has exactly a probability 
$\Gamma^{\rm E}_{lm}$ of being transmitted inside the object.

\section{Transfer matrix approach to high frequency resonances}\label{app:TransMatrix}
In the main text we have demonstrated the existence of oscillations and resonances in GFs and reflectivity. For low frequencies, these resonances can be interpreted as arising from the lowest-lying and long-lived quasi-normal modes, characterized by a very small imaginary part.
In this appendix we will discuss why even at high frequencies we have resonances and oscillations, 
which are evident in the reflectivity (left panels of Fig.~\ref{fig:SchwvsWorm} and Fig.~\ref{fig:QNMsvsresonance}) of the ECOs.  

In what follows, we will demonstrate the origin of high frequency oscillations and resonances through the transfer matrix method~\cite{Ianniccari:2024ysv}. 

\subsection{General framework}

In this method one starts in a region 
where the potential is negligible, and hence the master functions behave as a superposition of ingoing and outgoing plane waves. Since we are interested in asymptotically flat spacetimes, this is true at $r_*\to-\infty$ (near the ECO surface, or infinity in the mirror universe), $r_*\to\infty$ (infinity in our universe) and also in between the two potential barriers if they are sufficiently far apart in the tortoise coordinate. We can describe our master function via the following vector 
\begin{equation}
\Psi=\begin{pmatrix}
\psi_R \\
\psi_L
\end{pmatrix}\,
\end{equation}
where $\psi_{\rm R}$ is the right-going plane wave, while $\psi_{\rm L}$ the left-going one. 

In general, owing to linear nature of the perturbation equations, the master function at $r_{*}=\infty$ can be written in terms of the master function at $r_{*}=-\infty$, as,
\begin{equation}
\Psi(r_*=\infty)\,=\,\mathcal{M}\, \Psi(r_*=-\infty)\,
\end{equation}
where $\mathcal{M}$ is the transfer matrix. For scattering through a single potential, we may use the following sets of boundary conditions --- (a) for scattering of waves originating from $r_*=+\infty$ 
\begin{equation}\label{boundary_refl/trasmbis}
X^{\rm I}_{lm\omega}= 
\begin{cases}
e^{- {i} \omega r_*} \,\,\,r_*\to-\infty\\
\alpha^{\rm in}_{lm\omega} e^{- {i} \omega r_*}+ \alpha^{\rm out}_{lm\omega} e^{+ {i} \omega r_*} \,\,\,r_*\to+\infty 
\end{cases}\,,
\end{equation} 
while (b) for waves originating from $r_*=-\infty$, we have,
\begin{equation}\label{boundary_refl/trasmtris}
X^{\rm II}_{lm\omega}= 
\begin{cases}
\beta^{\rm in}_{lm\omega} e^{- {i} \omega r_*}+ \beta^{\rm out}_{lm\omega} e^{+ {i} \omega r_*}\,\,\,r_*\to-\infty\\
e^{{i} \omega r_*} \,\,\,r_*\to+\infty 
\end{cases}\,.
\end{equation}
Given the above complex amplitudes $\alpha_{lm\omega}^{\rm in/out}$ and $\beta_{lm\omega}^{\rm in/out}$, we can now introduce the following quantities,
\begin{align}\label{rtcoefficients}
R&={\alpha^{\rm out}_{lm\omega} \over  \alpha^{\rm in}_{lm\omega}}\,;
\quad
T={1 \over  \alpha^{\rm in}_{lm\omega}}\,,
\\
R'&={\beta^{\rm in}_{lm\omega} \over  \beta^{\rm out}_{lm\omega}}\,;
\quad
T'= {1 \over  \beta^{\rm out}_{lm\omega}}\,,
\end{align}
where, the terms without prime, namely $R,T$ (resp., with prime, namely $R',T'$) are related to reflection and transmission amplitudes of the potential barrier for plane waves coming from the right (resp., from the left). In terms of these quantities the transfer matrix $\mathcal{M}$ from left ($r_{*}\to-\infty$) to right ($r_{*}\to \infty$) becomes, 
\begin{equation}\label{schtranmat}
\mathcal{M}=\begin{pmatrix}
T'-{R R' \over T} & {R \over T} \\
-{R' \over T} & {1 \over T} \\
\end{pmatrix}\,.
\end{equation}
Due to the fact that the Wronskian associated with the differential equation describing the gravitational perturbation is a constant, it is possible to find out connection between the complex amplitudes $\alpha$ and $\beta$. 
In particular, considering the Wronskian made out of $(X^{\rm I}_{lm\omega},X^{\rm II}_{lm\omega})$ as well as $(X^{\textrm{I},*}_{lm\omega},X^{\rm II}_{lm\omega})$, we obtain,
\begin{equation}
\beta^{\rm out}_{lm\omega}=\alpha^{\rm in}_{lm\omega}~;
\qquad
\beta^{\rm in}_{lm\omega}=- \alpha^{\rm out,*}_{lm\omega}~.
\end{equation}
%
Note that the first condition ensures that the transmission amplitudes from the left and from the right coincide, i.e., $T=T'$.
The above can be easily generalized to the case of two potential barriers $V_{1}$ and $V_{2}$, separated by a distance $L$. In which case we can divide the full scattering matrix into three steps~\cite{Ianniccari:2024ysv}:
%
\begin{itemize}
    \item the scattering through the first barrier, described by $\mathcal{M}_1$;
    \item the dephasing in the passage from one barrier to another, described by a dephasing matrix $U(L)=\textrm{diag.}(e^{{i} \omega L},e^{-{i} \omega L})$, with $L$ being the distance between the two barriers;
    \item finally, the scattering through the second barrier, described by the transfer matrix $\mathcal{M}_2$.
\end{itemize}
Thus, except for an overall complex phase factor, we obtain,
\begin{equation}
\mathcal{M}=\mathcal{M}_2 U(L)\mathcal{M}_1\,,
\end{equation}
where $\mathcal{M}_{1,2}$ represent the transfer matrix for a single potential barrier. Given the individual transfer matrices and the de-phasing matrix, one can read off the total transmitivity of the two-barrier system to be,
\begin{equation}
{1 \over T}=-{R'_2 \over T_2}{R_1 \over T_1} e^{{i} \omega L}+{1 \over T_1 T_2}e^{-{i} \omega L}\,.
\end{equation}
Along similar lines, we can determine the left and the right reflectivities of the double barrier system from the above effective transfer matrix $\mathcal{M}$. The right and left reflection amplitudes can be explicitly written as
\begin{align}
\frac{R}{T}&=\frac{R_{2}}{T_{1}T_{2}}e^{-i\omega L}+\frac{R_{1}}{T_{1}}\left(T_{2}'-\frac{R_{2}R_{2}'}{T_{2}}\right)e^{i\omega L}\,,
\\
\frac{R'}{T}&=\frac{R_{1}'}{T_{1}T_{2}}e^{-i\omega L}+\frac{R_{2}'}{T_{2}}\left(T_{1}'-\frac{R_{1}R_{1}'}{T_{1}}\right)e^{i\omega L}\,.
\end{align}
Note that in general $R\neq R'$. The above expressions provide general results for transmission and reflection amplitudes, holding true for any well separated double barrier system. In the next subsections, we shall specialize to particular cases discussed in the main text.

\begin{figure*}[htbp!]
\centering
\includegraphics[width=0.95\textwidth]{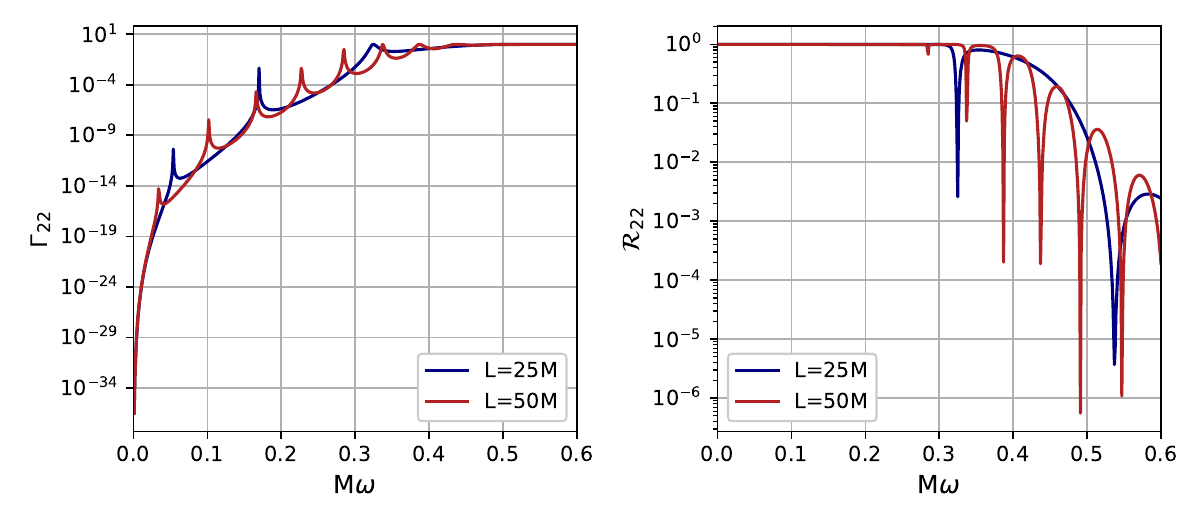}
\caption{GF and reflectivity computed via the transfer-matrix formalism for a double Schwarzschild potential with different dephasings in between the two barriers.}
\label{transfermatrix_variousL}
\end{figure*}

\subsection{Wormhole case}\label{app:TransMatrixWorm}

Specializing to the case of the Schwarzschild wormhole, where the two barriers are mirror symmetric of one another, in the right universe, for the transfer matrix $\mathcal{M}_{2}$, we can directly use the expressions in Eq.~\eqref{rtcoefficients}. In the left universe, on the other hand, we need to exchange the primed and the unprimed quantities, since the potential is mirror symmetric about $r_{*}=0$. This description implies that the total transmission ($T_{\rm w}$) and reflection ($R_{\rm w}$) amplitudes for the wormhole to be
\begin{eqnarray}    
T_{\rm w}&=& {1 \over   \left(\alpha^{\rm in}_{lm \omega}\right)^2e^{-{i} \omega L}-\left(\alpha^{\rm out,*}_{lm\omega}\right)^2 e^{{i} \omega L}}~,\label{t_transmatrix}\\
\frac{R_{\rm w}}{T_{\rm w}}&=&\alpha_{lm\omega}^{\rm in}\alpha_{lm\omega}^{\textrm{out}}e^{-{i} \omega L}
-\alpha_{lm\omega}^{\textrm{in},*}\alpha_{lm\omega}^{\textrm{out},*}e^{i\omega L}~,\label{r_transmatrix}
\end{eqnarray}
where, $\alpha^{\rm in/out}_{lm\omega}$ are the standard ingoing and outgoing amplitudes of the Schwarzschild BH satisfying $1+|\alpha_{lm\omega}^{\rm out}|^{2}=|\alpha_{lm\omega}^{\rm in}|^{2}$. Clearly, the GF for a wormhole is given by $\Gamma=|T_{\rm w}|^2$, as reported in the main text. One can also verify that $|T_{\rm w}|^{2}+|R_{\rm w}|^{2}=1$. 

Given the above general formalism for GF of wormholes based on transfer matrix method, let us try to understand the role of the dephasing in the computation of the GFs and also the reflectivity. This has been clearly depicted in Fig.~\ref{transfermatrix_variousL}, where we have presented both the GF and the reflectivity for the $l=2=m$ mode against the frequency for different choices of $L$. As evident, with increasing the value of $L$, the number of oscillation increases. This can be compared with the result we had obtained in Eq.~\eqref{t_transmatrix}, with increase of the length $L$, the dephasing caused by the $e^{i\omega L}$ term increases. Therefore, it is clear that the dephasing, due to increase of the distance between the two barriers, indeed leads to an increase in the number of oscillations in GF and reflectivity.

\begin{figure}[]
    \centering
\includegraphics[width=0.5\textwidth]{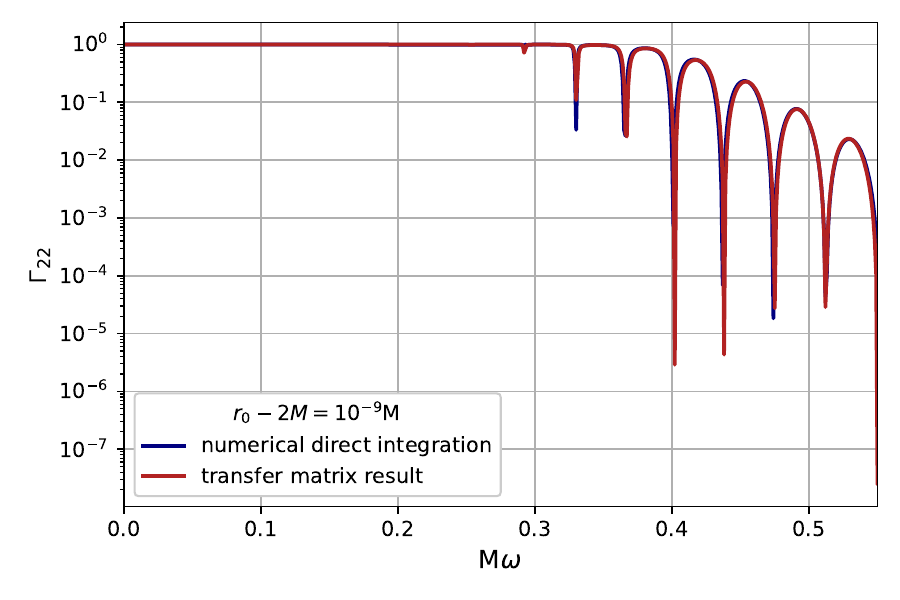}
    \caption{Comparison between the GF of a wormhole computed using Eq.~\eqref{t_transmatrix} and the exact numerical result for the representative case $r_0=(2+10^{-9})M$. The distance used for the reconstruction is $L=10\omega+67$, which models the distance in tortoise coordinate between the barriers as a function of the frequency at first order. Clearly, such a linear interpolation can work only as long as $M\omega<1$, as it is in our case.}
    \label{reflreconstruction}
\end{figure}

It is important to make a significant clarification at this stage. Indeed, not all plane waves are going to ``see" the same distance between the two barriers. In other words, the distance between the two barriers can be expressed as $L=L(\omega)$, with a non-trivial dependence of the barrier separation on the frequency $\omega$. This is due to the fact that the barriers are not perfect step functions. In Fig.~\ref{reflreconstruction} we use a linear interpolation to include such a dependence of $L$ on $\omega$ at least at first order (one could improve it adding more orders). Even including only a first order dependence on $\omega$ for $L$ the reconstruction with the transfer matrix formalism works very well.
\subsection{Partially absorbing ECOs} 

In the previous appendix, we have developed the transfer matrix approach and have applied it to a Schwarzschild ECO. In this appendix, we wish to apply it to reflective compact objects, which consist of a photon sphere and a reflective surface near the horizon. We start with the case of Boltzmann ECO as a warm up example and then provide the GF and reflectivity for generic ECOs. 

\subsubsection{Boltzmann ECOs}\label{appendixECOBoltz}

As a first example of a reflective compact object, we start with the case of Boltzmann ECOs, whose reflectivity is frequency dependent and is simply given by $\exp(-|\omega|/T_{\rm H})$. We can apply the transfer matrix formalism developed in Appendix~\ref{app:TransMatrixWorm}. But in this case, the initial state is not at $r_{*}\to-\infty$, rather at the ECO radius $r_{*}^{0}$, such that the decomposition of the perturbation variable in left and right moving waves, read
\begin{equation}
\Psi(r_0)=\begin{pmatrix}
e^{-\abs{\omega}/T_H}e^{-i\omega r_{*}^{0}} \\ e^{i\omega r_{*}^{0}}
\end{pmatrix}\,.
\end{equation}
Here, we have used Eq. \eqref{ECObcs}, for describing the perturbation variable near the ECO surface. Thus the perturbation at asymptotic infinity takes the following form:
\begin{equation}
\Psi(\infty)=\mathcal{M}U(-r_{*}^{0})\Psi(r_0)\,,
\end{equation}
with $\mathcal{M}$ being the standard Schwarzschild transmittivityBH transfer matrix, presented in Eq.~\eqref{schtranmat}, and $U(-r_{*}^{0})$ is the dephasing matrix discussed in Appendix~\ref{app:TransMatrixWorm}. The above relation further assumes that the photon sphere is located at $r_{*}=0$. This implies the following expression for the GF (see Eq.~\eqref{ECOgfs} introduced above):
\begin{equation}
{1 \over  \Gamma^{\rm B}_{lm}}=\left|\alpha_{lm\omega}^{\rm in}+\alpha_{lm\omega}^{\rm out,*} e^{-4i \omega r_{*}^{0}}e^{-\abs{\omega}/T_H}\right|^{2}\left(1-e^{-2|\omega|/T_{\rm H}}\right)~,
\end{equation}
with $\alpha_{lm\omega}^{\rm in/out}$ being the standard ingoing and outgoing Schwarzschild amplitudes for plane waves. Similarly, the reflectivity for the Boltzmann ECO takes the following form, 
\begin{align}\label{reflectivity_boltzmann}
\mathcal{R}^{\rm B}_{lm}&=\left|\frac{\alpha_{lm\omega}^{\rm out}+\alpha_{lm\omega}^{\textrm{in},*}e^{-\abs{\omega}/T_H}e^{-4i\omega r_{*}^{0}}}{\alpha_{lm\omega}^{\rm in}+\alpha_{lm\omega}^{\textrm{out},*}e^{-\abs{\omega}/T_H}e^{-4i\omega r_{*}^{0}}}\right|^{2}~.
\end{align}
Here, we have used the result, $|\alpha_{lm\omega}^{\rm in}|^{2}=1+|\alpha_{lm\omega}^{\rm out}|^{2}$, using which one can show that $\mathcal{R}^{\rm B}_{lm}+\Gamma^{\rm B}_{lm}=1$. Thus given these reflection and transmission amplitudes associated with the BH potential, the GF as well as the reflectivity for Boltzman ECO can be uniquely determined. 

\subsubsection{GFs and reflectivity of ECOs: General results}\label{appendixECOconst}

The above result for the Boltzman ECO can be easily generalized to any ECOs with reflectivity $R_{\rm ECO}$, defined through the perturbation near the ECO surface, such that,
\begin{equation}
\Psi(r_0)=\begin{pmatrix}
R_{\rm ECO}e^{-i\omega r_{*}^{0}} \\ e^{i\omega r_{*}^{0}}
\end{pmatrix}\,.
\end{equation}
This definition follows from the behaviour of the master function near the ECO surface, as presented in Eq.~\eqref{ECObcs}. Thus, following the steps of the previous section, assuming that the exterior spacetime is given by the Schwarzschild metric, the GF for any ECO can be expressed as\footnote{Note that the Wronskian condition: $\beta^{\rm out}_{lm\omega}=\alpha_{lm\omega}^{\rm in}$ automatically guarantee that the left and right transmission amplitudes are identical. The choice that right and left reflection amplitudes are also identical yields the following relation: $\alpha_{lm\omega}^{\rm out}=-\alpha_{lm\omega}^{\textrm{out},*}$, which we will not use here.},
\begin{align}
\frac{1}{\Gamma^{\rm E}_{lm}}&=\left|\frac{1}{T_{\rm BH}}-\frac{R_{\rm ECO}R'_{\rm BH}e^{-4i\omega r_{*}^{0}}}{T_{\rm BH}}\right|^{2}\left(1-|R_{\rm ECO}|^{2}\right)
\nonumber
\\
&=\left|\alpha_{lm\omega}^{\rm in}+\alpha_{lm\omega}^{\rm out,*}R_{\rm ECO}e^{-4i\omega r_{*}^{0}}\right|^{2}\left(1-|R_{\rm ECO}|^{2}\right)~.
\end{align}
In arriving at the above expression, we have assumed that the $r_{*}=0$ corresponds to the location of the photon sphere. The reflectivity, on the other hand, can be expressed as,
\begin{align}\label{reflectivity}
\mathcal{R}^{\rm E}_{lm}&=\left|\frac{\frac{R_{\rm BH}}{T_{\rm BH}}+\left(T_{\rm BH}-\frac{R_{\rm BH}R_{\rm BH}'}{T_{\rm BH}}\right)R_{\rm ECO}e^{-4i\omega r_{*}^{0}}}{\frac{1}{T_{\rm BH}}-\frac{R_{\rm ECO}R'_{\rm BH}e^{-4i\omega r_{*}^{0}}}{T_{\rm BH}}}\right|^{2}
\nonumber
\\
&=\left|R_{\rm BH}+\frac{T_{\rm BH}^{2}R_{\rm ECO}e^{-4i\omega r_{*}^{0}}}{1-R'_{\rm BH}R_{\rm ECO}e^{-4i\omega r_{*}^{0}}}\right|^{2}~.
\end{align}
The above expression can also be understood as follows: consider a wave originating from infinity, whose transmission amplitude through the BH potential barrier is $T_{\rm BH}$ (which is the same from the left and from the right) and the reflection amplitude of the potential barrier is $R_{\rm BH}$ from the right and $R'_{\rm BH}$ from the left. There will be an outgoing amplitude at infinity, due to the infinite series of transmission and reflections from the surface of the ECO as well as the photon sphere. Denoting reflection amplitude of the ECO by $R_{\rm ECO}e^{-2i\omega r_{*}^{0}}$ (to take care of the boundary condition, as discussed in Eq.~\eqref{ECObcs} of the main text), we obtain
\begin{align}
\nonumber \vphantom{\sum} A^{\rm out}&= A^{\rm out,BH}
+R_{\rm ECO}T_{\rm BH}e^{-4i\omega r_{*}^{0}}
\\
&\vphantom{\sum}\nonumber+R_{\rm ECO}^2 T_{\rm BH} R'_{\rm BH}e^{-8i\omega r_{*}^{0}}+\cdots
\\
&=A^{\rm out,BH}+T_{\rm BH}R_{\rm ECO}e^{-4i\omega r_{*}^{0}}
\nonumber
\\
&\qquad \times \sum_{n=0}^{\infty} \left(R_{\rm ECO}R'_{\rm BH}e^{-4i\omega r_{*}^{0}}\right)^n\,,
\end{align}
where, we have used the result that each reflection from the ECO surface generates a phase factor of $e^{-2i\omega r_{*}^{0}}$, as the wave traverses the cavity twice. In addition, the above result uses the following identity: $A_{\rm in}=T_{\rm BH}^{-1}$. Performing the infinite summation in the last step, we obtain the reflectivity to be identical to Eq.~\eqref{reflectivity}. Then the identity: $\Gamma^{\rm E}_{lm}=(1-\mathcal{R}^{\rm E}_{lm})$, provides the GF of ECO.
This further depicts the correctness of the GF and reflectivity obtained through the transfer matrix method for generic ECOs. The above expression have been numerically verified to reconstruct the total reflectivity at all frequencies when the ECO radius approaches the would be horizon. 
Moreover, the mapping between the reflectivity of the ECO and the wormhole can be obtained by the following relation: $R_{\rm ECO}=R'_{\rm BH}$. Note that our result differs from \cite{Mark:2017dnq} in two aspects --- (a) We do not have the phase factor $\exp(-2i\omega r_{*}^{0})$, as in the transfer matrix approach this is already taken care of by the translation matrix $U(-r_{*}^{0})$; (b) the reflective amplitude of the ECO, $R_{\rm ECO}$ should be related to the left side reflectivity of the BH potential, which follows from the fact that the potential on the other side of the throat is a mirror symmetry of the potential on our side of the universe. Note that, this choice ensures that $\mathcal{R}_{lm}^{\rm w}=\mathcal{R}_{lm}^{\rm E}$, i.e., the two reflectivities are identical, as well as, $\Gamma_{lm}^{\rm w}=\Gamma_{lm}^{\rm E}$. We would like to emphasize that the above equality requires the extra factor of $(1-|R_{\rm ECO}|^{2})$ in the definition of the GF as advocated in Appendix \ref{app:GFSforECOs}.

\begin{figure*}[htbp!]
\centering
\includegraphics[width=\textwidth]{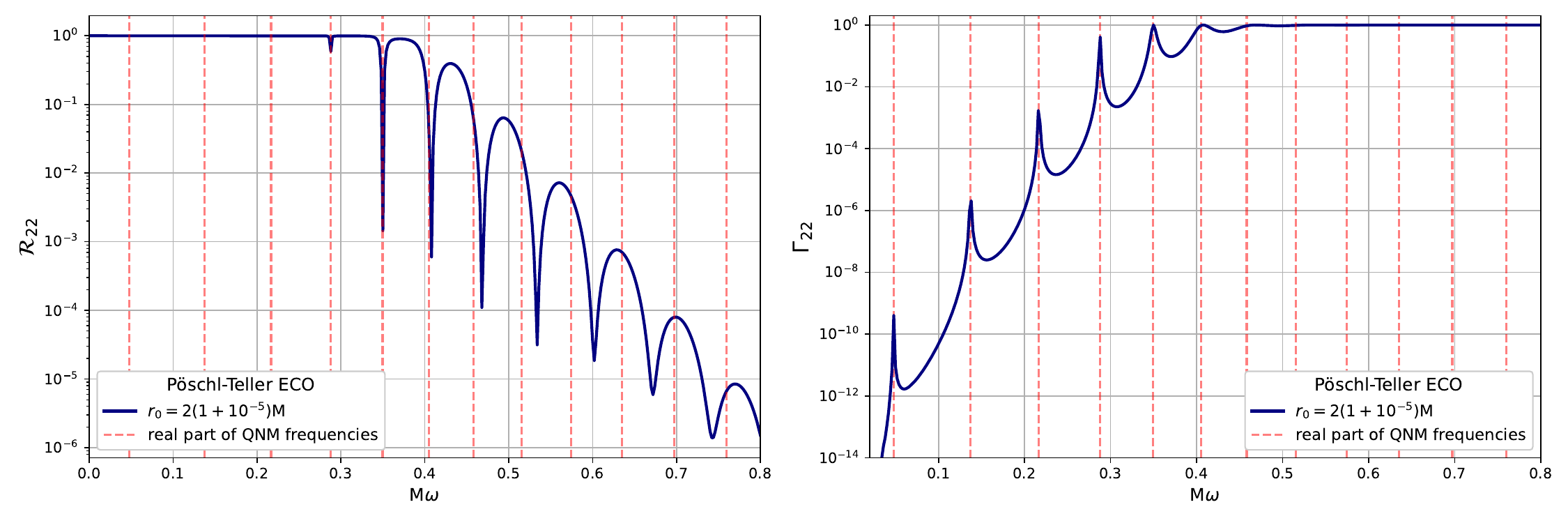}
\caption{The reflectivity $\mathcal{R}_{22}$ and the GF $\Gamma_{22}$ are being presented against the dimensionless frequency $M\omega$ for the dominant $l=2=m$ modes associated with the scattering of gravitational perturbation from an ECO. For a comparison with the Schwarzschild wormhole case, presented in Fig.~\ref{fig:QNMsvsresonance}, we have considered the ECO with Pöschl–Teller-like surface reflectivity and with radius $r_0/2M=(1+10^{-5})$.}
\label{fig:PTECO}
\end{figure*}

To show that above correspondence between wormhole and ECO explicitly, we have considered an ECO with a Pöschl–Teller reflectivity, so that it can mimic a Schwarzschild wormhole. This is because, the Pöschl–Teller potential can effectively reproduce the photon sphere potential of Schwarzschild metric and is more easily tractable analytically. Following which, in Fig.~\ref{fig:PTECO} we have shown the behavior of reflectivity and transmitivity for an ECO with such a Pöschl–Teller reflectivity, which can be compared with Fig.~\ref{fig:QNMsvsresonance} for a Schwarzschild wormhole. As evident, both the figures depict that one can get the correct wormhole phenomenology for reflectivity and GFs from an ECO with an appropriate choice of the reflectivity of the ECO. 

\section{Small-frequency analytical approximation to the GF of ECO}\label{analyticgreybody}

In this appendix, we will provide 
an analytical approximation for the GF of a Schwarzschild ECO using scalar perturbation. The technique presented here will be based on the matched asymptotic expansion method \cite{Starobinsky:1973aij}, and can be generalized to generic spin perturbation in a straightforward manner. (See Ref.~\cite{Malik:2024wvs} for a recent analysis in the wormhole case).
This method provides a good approximation for the scattering quantities in the low frequency regime ($M\omega \ll1$) and we will apply it for different ECOs. We start by decomposing the scalar field $\Phi$ into spherical harmonic basis,
\begin{align}
\Phi=\int dt \sum_{l m}e^{-i\omega t}\phi_{l m}(r)Y_{l m}(\theta,\phi)~.
\end{align}
Given the radial function $\phi_{lm}(r)$, we can introduce another function $R_{lm}=r^{-1}\phi_{lm}$, which satisfies the following differential equation,
\begin{align}
\label{EQgeneralMM}
\nonumber f(r) \frac{d}{dr} &\left(f(r) r^2 \frac{dR_{l m}}{dr}\right) \\&\,\,\,\,\,\,\,\,\,\,\,+ \left[r^2 \omega^2 - f(r) \{l(l + 1)  \}\right] R_{l m} = 0\,.
\end{align}
The subsequent steps of this method consists of analytically solving the previous equation in the far region and in the near-surface region of the ECO. Then we will match the two solutions in the intermediate region and obtain the GF. In the following we will set $M=1$ so that all of our quantities become dimensionless.

\paragraph*{Solution in far region}--- In the far region limit, which corresponds to $r\gg M$, the perturbation equation for the radial sector, as presented in Eq.~\eqref{EQgeneralMM}, becomes
\begin{equation}
r^2\frac{d^2 R^{\rm F}_{lm}}{dr^2}+2r\frac{dR_{lm}^{\rm F}}{dr} + \left[r^2 \omega^2 - l(l + 1)\right]R_{lm}^{\rm F}=0~.
\end{equation}
General solution of this equations can be expressed in term of Bessel functions of first kind,
\begin{equation}
R_{lm}^{\rm F} = \frac{1}{\sqrt{r}} \left[\alpha J_{l + \frac{1}{2}}(\omega r) + \beta J_{-l - \frac{1}{2}}(\omega r)\right]\,.
\end{equation}
By using the asymptotic properties of the Bessel functions, asymptotically ($r\to\infty$), we have,
\begin{equation}
J_{l+\frac{1}{2}}(\omega r) \sim \sqrt{\frac{2}{\pi \omega r}} \cos \left[\omega r - \frac{\pi}{2} \left(l + \frac{1}{2}\right) - \frac{\pi}{4}\right]\,.
\end{equation}
Consequently, the radial part of the scalar perturbation reduces to,
\begin{equation}
R_{lm}^{\rm F}(r \to \infty) \sim \sqrt{\frac{1}{ 2\pi \omega}} \left(\alpha' e^{i\omega r} + \beta' e^{-i\omega r}\right)\,,
\end{equation}
with, the coefficients $\alpha'$ and $\beta'$ being expressed as a linear combination of the arbitrary constants $\alpha$ and $\beta$,
\begin{equation}
\alpha'=\beta e^{(i\pi l/2)}-i\alpha e^{-(i\pi l/2)}~; 
\quad 
\beta'=i\alpha e^{(i\pi l/2)}+\beta e^{-(i\pi l/2)}\,.
\end{equation}
Since, the above computation holds for $(r/M)\gg 1$, it follows that $r_{*}\approx r$, and hence the ingoing and outgoing amplitudes at infinity are given by,
\begin{equation}
A^{\rm out}_{\rm lm \omega}=\sqrt{\frac{1}{ 2\pi \omega}} \alpha'\,\,\,\,\,\,\,A^{\rm in}_{\rm lm \omega}=\sqrt{\frac{1}{ 2\pi \omega}} \beta'\,.
\end{equation}
Therefore, it follows that the complex reflection amplitude of the scalar perturbation from the ECO will be,
\begin{equation}\label{ReflAnalytic}
R_{\rm lm}(\omega)=-\frac{\alpha +i\beta e^{i\pi l}}{\alpha  e^{i\pi l}-i\beta }~.
\end{equation}
Now, considering the intermediate region, using the properties of the Bessel functions for $r\ll 1$, we have,
\begin{align}
\nonumber R_{lm}^{\rm F\,(int)}(r)&\sim \alpha \left(\frac{\omega }{2}\right)^{l+\frac{1}{2}}\frac{r^{l}}{\Gamma\left(l+\frac{3}{2}\right)} 
\\
&\,\,\,\,\,\,\,\,\,+ \beta \left(\frac{\omega }{2}\right)^{-l-\frac{1}{2}} 
\frac{r^{-l-1} }{\Gamma\left(\frac{1}{2}-l\right)}\,.
\end{align}
Thus the far zone solution has two branches in the intermediate region, one growing as $r^{l}$, and the other decaying as $r^{-l-1}$. We now turn our attention to the region near the ECO surface.

\paragraph*{Near zone solution}---In the near zone, it is useful to define $x\equiv (r/2)-1$, so that the horizon is mapped to $x=0$. Then the radial perturbation equation, as presented in Eq.~\eqref{EQgeneralMM} becomes
\begin{align}
\nonumber x&(1+x)\frac{d}{dx}\left[x(1+x)\frac{dR_{lm}^{\rm N}}{dx}\right] 
\\
&
\,\,\,\,\,+ (1+x)\left[\bar{\omega}^2(1+x)^3-xl(l+1)\right]R_{lm}^{\rm N}=0\,,
\end{align}
with $\bar{\omega}\equiv 2\omega$. With the following redefinition of the radial perturbation variable: $R_{lm}^{\rm N}=(1+x)^{-i\bar{\omega}}x^{i\bar{\omega}}F$, it follows that $F$ satisfies the hypergeometric differential equation, and hence, the radial perturbation equation in the near zone reads,
\begin{align}
\nonumber & R_{lm}^{\rm N}(x)=c_{1}(1+x)^{-i\bar{\omega}} x^{i\bar{\omega}}{}_2F_1(-l, l+1, 1+2i\bar{\omega}, -x) 
\\
&+c_{2} (1+x)^{-i\bar{\omega}} x^{-i\bar{\omega}}
\nonumber
\\
&\qquad \times {}_2F_1(-l-2i\bar{\omega}, l+1-2i\bar{\omega}, 1-2i\bar{\omega},-x)~.
\end{align}
Considering the properties of the hypergeometric function, namely ${}_2F_1(a, b, c, 0)=1$, it follows that near the surface of the ECO (obtained by the $x \to 0$ limit), we have
\begin{equation}
R_{lm}^{\rm N}(x) \sim 
c_1 e^{-i\omega r_*} + c_2 e^{i\omega r_*}\,,
\end{equation}
hence the choice of the arbitrary constants $c_{1}$ and $c_{2}$ will set the boundary conditions at the near-horizon regime, which is purely transmission for a BH, or, both reflection and transmission at the effective radius for ECOs. Thus for a BH, we must set $c_{2}=0$, while for an ECO, both of these constants are non-vanishing, with $R_{\rm ECO}=(c_{2}/c_{1})$. Moreover, the near-zone radial function can be determined in the intermediate region, by considering $x \to \infty$ limit of $R_{lm}^{\rm N}$. To obtain this limit, we provide the asymptotic behaviour of the hypergeometric function, which reads,
\begin{equation}
{}_2F_1(a, b; c; z) = \frac{\Gamma(c)\Gamma(b-a)}{\Gamma(b)\Gamma(c-a)} (-z)^{-a} + (a \leftrightarrow b)\,,
\end{equation}
and hence for $r\gg1$, i.e., in the intermediate region, we have 
\begin{align}
&R_{lm}^{\rm N\,(int)}(r)
\nonumber
\\
&\sim \left({r \over 2}\right)^{l} \frac{\Gamma(2l+1)}{\Gamma(l+1)} \left[\frac{\Gamma(1+2i\bar{\omega})}{\Gamma(1+2i\bar{\omega}+l)} c_1+\frac{\Gamma(1-2i\bar{\omega})}{\Gamma(1-2i\bar{\omega}+l)} c_2 \right]
\nonumber
\\
&\quad+\left({r \over 2}\right)^{-l-1} \frac{\Gamma(-2l-1)}{\Gamma(-l)} \left[\frac{\Gamma(1+2i\bar{\omega})}{\Gamma(2i\bar{\omega}-l)} c_1 +\frac{\Gamma(1-2i\bar{\omega})}{\Gamma(-2i\bar{\omega}-l)} c_2 \right]\,.
\nonumber
\end{align}
This provides the contribution to the near-zone region from the radial part of scalar perturbation.

\paragraph*{Matching}--- We now have to match the radial perturbation in the far and the near zone, as derived above, in the intermediate region, i.e.,
\begin{equation}
R_{lm}^{\rm N\,(int)}(r)=R_{lm}^{\rm F\,(int)}(r)\,.
\end{equation}
Thus equating the coefficients of $r^{l}$ and $r^{-l-1}$ in both of these expressions, we can extract the coeffients $\alpha$ and $\beta$ in the far zone solution: 
\begin{align}
\nonumber \,\,\,\,\,\alpha&=\frac{ \Gamma\left(l+\frac{3}{2}\right)\Gamma(2l+1) \sqrt{2}}{\Gamma(l+1) }  \\& \times \nonumber\left[\frac{\Gamma(1+2i\bar{\omega})}{\Gamma(1+2i\bar{\omega}+l)} c_1 +\frac{\Gamma(1-2i\bar{\omega})}{\Gamma(1-2i\bar{\omega}+l)} c_2 \right] \left(\omega \right)^{-l-\frac{1}{2}}\,,
\\
\nonumber \,\,\,\,\,\beta&=\frac{\Gamma(-2l-1)\Gamma\left(\frac{1}{2}-l\right)}{\Gamma(-l) \sqrt{2}} \\&\times  \left[\frac{\Gamma(1+2i\bar{\omega})}{\Gamma(2i\bar{\omega}-l) } c_1 +\frac{\Gamma(1-2i\bar{\omega})}{\Gamma(-2i\bar{\omega}-l)} c_2 \right] \left(\omega \right)^{l+\frac{1}{2}} \,.
\end{align}
It might appear that the above expressions are divergent in the $l\to\textrm{integer}$ limit. However, one can use the following identity involving $\Gamma$ functions to show the finiteness of the results,
\begin{equation}
\frac{\Gamma(-2l-1)\Gamma\left(\frac{1}{2}-l\right)}{\Gamma(-l)}= -\pi \Gamma(l + \frac{1}{2}) {\Gamma(l+1) \over \Gamma(2l+2)}\,.
\end{equation}
Using this identity, as well as the fact that the ratio $(c_{2}/c_{1})$ is related to the reflectivity of the compact object, we can obtain a closed form expression for the reflection and transmission amplitudes using Eq.~\eqref{ReflAnalytic}. This means that we are able to compute the reflectivity and GFs for various boundary conditions. These are summarized in Table \ref{tab:example}.

\renewcommand{\arraystretch}{1.5}
\begin{table}[h!]
\centering
\begin{tabular}{|c|c|c|}
\hline
\textbf{Gravitational Object} & \(\mathbf{c_1}\) & \(\mathbf{c_2}\) \\ \hline
Schwarzschild BH & \(0\) & \(1\) \\ \hline
Schwarzschild-like wormhole & \(R'^{\rm BH}_{\rm lm}(\omega)e^{-4i\omega r_*(r_0)}\) & \(1\) \\ \hline
ECO & \(R^{\rm ECO}_{\rm lm}(\omega)e^{-2i \omega r_*(r_0)}\) & \(1\) \\ \hline
\end{tabular}
\caption{Boundary conditions for different compact objects have been presented. In the wormhole case, $r_{0}$ corresponds to the throat position, while in the ECO case $r_{0}$ represents the radius of the object. The ECO boundary condition accounts for any ECO model given the reflectivity at $r_0$.}
\label{tab:example}
\end{table}

\begin{figure}
\centering
\includegraphics[width=\linewidth]{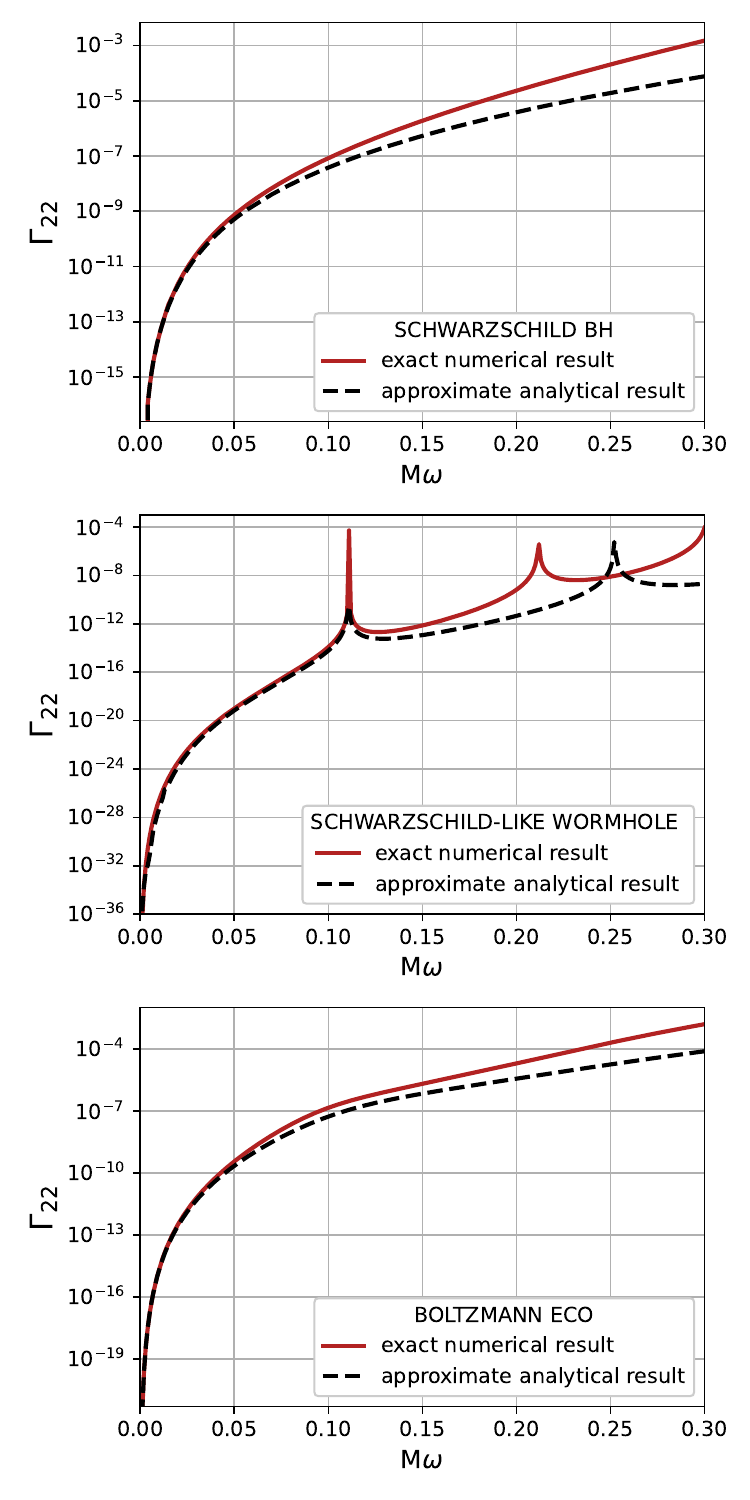}
\caption{GF for the $s=0$ perturbation for various compact objects. In the wormhole case, we have taken, $r_0=(2+10^{-6})M$, while for the Boltzmann ECO case, we have taken $r_0=(2+10^{-4})M$. As evident, in all of these cases, the analytical result matches with the numerical result at low frequencies, while at high frequencies there are differences.}
\label{ECOmm}
\end{figure}
Following our analytic computation, in Fig.~\ref{ECOmm} we compare the exact numerical results with the analytical ones, obtained here, for various compact objects of interest. In all of the cases the approximation is in good accordance with the correct result at low frequencies, while at larger ones it starts to be less compatible.

\bibliography{biblio}
\end{document}